\documentclass[sigconf,nonacm]{acmart}
\AtBeginDocument{%
  }

\usepackage{hyperref}
\usepackage{url}
\usepackage{amsmath}
\usepackage{color}
\usepackage{appendix}
\usepackage{booktabs}
\usepackage{subcaption}
\usepackage{booktabs}
\usepackage{graphicx}
\usepackage[normalem]{ulem} 
\usepackage{mathtools}
\usepackage{xcolor}
\usepackage{glossaries}
\usepackage{multirow}

\usepackage{algorithm}
\usepackage[noend]{algpseudocode}
\newcommand{\LL}[1]{{\color{red}[LL: #1]}}
\newcommand{\ml}[1]{{\color{blue}[ML: #1]}}

\newcommand{\unsafe}{\mathrm{u}}
\newcommand{\safe}{\mathrm{s}}
\newcommand{\reach}{\mathrm{r}}
\newcommand{\initial}{\mathrm{0}}

\newcommand{\pw}{\mathbf{w}}
\newcommand{\px}{\mathbf{x}}

\newcommand{\pX}{\mathbf{x}}

\newcommand{\reals}{\mathbb{R}}
\newcommand{\naturals}{\mathbb{N}}

\newcommand{\N}{\mathcal{N}}
\newcommand{\hor}{H}
\newcommand{\RA}{\mathcal{R}}
\newcommand{\naturalszero}{\mathbb{N}_{\geq 0}}

\newcommand{\nx}{\mathbf{x}'}

\newcommand{\Pfin}{P_{\reach}(X_{\reach}, X_u, X_0, \hor, \pi)}
\newcommand{\Pfinopt}{P_{\reach}(X_{\reach}, X_u, X_0, \hor, \pi^{\ast})}
\newcommand{\Pinf}{P_{\reach}(X_{\reach}, X_u, X_0, \pi)}

\newtheorem{theorem}{Theorem}
\newtheorem{problem}{Problem}
\newtheorem{remark}{Remark}
\newtheorem{definition}{Definition}

\newtheorem{corollary}{Corollary}

\newacronym{lp}{LP}{Linear Programming}
\newacronym{gd}{GD}{Gradient Descent}
\newacronym{pwc}{PWC}{Piecewise Constant}
\newacronym{pwa}{PWA}{Piecewise Affine}
\newacronym{sos}{SOS}{Sum-of-Squares}
\newacronym{sdp}{SDP}{Semi-Definite Programming}
\newacronym{cbfs}{CBFs}{Control Barrier Functions}
\newacronym{scbf}{s-CBFs}{Stochastic Control Barrier Functions}
\newacronym{sbfs}{SBFs}{Stochastic Barrier Functions}
\newacronym{pscbf}{PS-CBFs}{Piecewise Stochastic Control Barrier Functions}
\newacronym{cpscbf}{cPS-CBFs}{Constant Piecewise Stochastic Control Barrier Functions}
\newacronym{ibp}{IBP}{Interval Bound Propagation}
\newacronym{imdp}{IMDP}{Interval Markov Decision Process}
\newacronym{idlp}{IDLP}{Iterative Dual Linear Program}
\newacronym{rgd}{RGD}{Riemannian Gradient Descent}
\newacronym{pwf}{PWF}{Piecewise Feedback}
\newacronym{dp}{DP}{Dynamic Programming}

\newif\ifarxiv

\arxivtrue

\ifarxiv
\setcopyright{none}
\acmDOI{}
\acmISBN{}
\acmConference{}{}{}
\acmYear{}
\fi

\begin{document}

\title{
Scalable Synthesis of Time-Varying Stochastic Control Certificates via Sequential Optimization
}

\ifarxiv
    \author{Rayan Mazouz}
    \email{rayan.mazouz@colorado.edu}
    \authornotemark[1]
    \affiliation{%
      \institution{University of Colorado Boulder}
      \city{Boulder}
      \state{Colorado}
      \country{USA}
    }
    
    \author{Luca Laurenti}
    \email{l.laurenti@tudelft.nl}
    \affiliation{%
      \institution{Delft University of Technology}
      \city{Delft}
      \country{Netherlands}
    }

        \author{Morteza Lahijanian}
    \email{morteza.lahijanian@colorado.edu}
    \affiliation{%
      \institution{University of Colorado Boulder}
      \city{Boulder}
      \state{Colorado}
      \country{USA}
    }

\else
    \author{Anonymous Author(s)}
    \email{}
    \affiliation{%
      \institution{}
      \city{}
      \state{}
      \country{}
    }
\fi

\ifarxiv
\renewcommand{\shortauthors}{Mazouz et al.}
\fi
\begin{abstract}
Joint controller-certificate synthesis is a powerful framework for establishing probabilistic safety and reach-avoid guarantees for stochastic dynamical systems. 
Recent time-varying certificate formulations yield tighter probability bounds than classical time-invariant approaches but require large-scale optimization, particularly when controllers and certificates are synthesized jointly.
In this paper, we introduce a scalable sequential framework for joint controller-certificate synthesis that preserves formal probabilistic guarantees. We focus on reach-avoid properties and show that the framework naturally extends to safety properties. By decomposing the original optimization problem into a sequence of smaller sum-of-squares programs, the proposed framework significantly improves computational scalability. However, sequential decomposition introduces sub-optimality because each optimization step lacks complete information about the remaining synthesis horizon. To mitigate this, we propose two approaches: (i) uniform subspace regularization and (ii) initial-set tracking through nominal dynamics propagation. Experiments across a range of benchmarks demonstrate that the proposed framework substantially improves scalability over existing formulations while achieving competitive certified probability bounds for both reach-avoid and safety properties.
\end{abstract}



\maketitle

\section{Introduction}


Ensuring the safe operation of autonomous systems in \emph{safety-critical} applications is a central objective in modern control and robotics~\cite{guiochet2017safety}.
Real-world systems are subject to uncertainty and stochastic disturbances, making their behavior difficult to predict. Consequently, it is necessary to rigorously verify system behavior, and synthesize controllers that provide \emph{formal} probabilistic guarantees. Reach-avoid control-certificates~\cite{mazouz2026stochastic, mazouz2026timevarying} provide a powerful framework for this task: a certificate satisfying a collection of functional inequalities yields a probabilistic lower bound on the reach-avoid probability~\cite{kushner1967stochastic}, and can be synthesized jointly with a controller, e.g., via convex \gls{sos} programming~\cite{legat2017sos, mazouz2022safety}. However, classical formulations rely on a single certificate enforced uniformly across a horizon, limiting their ability to capture the sequential structure of stochastic dynamical evolution. Recently introduced \emph{time-varying} reach-avoid certificates~\cite{mazouz2026timevarying, mazouz2026stochastic} address this conservatism through a \gls{dp} characterization of (in)finite-horizon reach-avoid probabilities, yielding tighter probabilistic guarantees. However, the optimization problem is formulated as a single \emph{oneshot} \gls{sdp} that synthesizes all stage-wise certificates simultaneously, limiting scalability. In this work, we address this challenge through a \emph{sequential} certificate synthesis framework that enables efficient joint controller-certificate synthesis.

Existing time-varying reach-avoid certificate formulations couple stage-wise certificates through backward expectation constraints and a terminal initial condition~\cite{oumer2024safety, mazouz2026stochastic, mazouz2026timevarying}. Consequently, controller-certificate synthesis is performed in a single optimization problem over the entire horizon. A natural question is whether the certificates can instead be constructed sequentially, treating previously synthesized certificates as fixed and solving a sequence of smaller optimization problems. The difficulty is that the initial set condition, which determines the final lower bound on the reach-avoid probability, is only enforced at the final stage of the recursion and is therefore unavailable during intermediate stages. Sequential synthesis therefore requires surrogate objectives that capture the influence of this final condition throughout the certificate construction process. The reduced computational cost of sequential synthesis, however, permits higher-degree certificate templates, which can compensate for the loss in optimality and recover the reach-avoid probability bounds of the oneshot formulation.

In this paper, we develop a sequential framework for jointly synthesizing time-varying reach-avoid controllers and certificates for discrete-time nonlinear stochastic systems over continuous state spaces. The proposed formulation follows a backward DP structure while decomposing the large \gls{sdp} into a sequence of smaller \gls{sos} programs. To approximate the influence of the initial-set condition, we introduce two surrogate methods. The first uses Monte Carlo simulations to propagate a finite set of samples from the initial set and regularizes the certificate over the resulting sampled states at each time step.
The second draws uniformly distributed samples across the state space and regularizes the certificate at these sampled states, promoting uniformly large certificate values throughout the state space. We evaluate the formulations on nonlinear systems ranging from one to eight dimensions and compare them against the time-invariant~\cite{mazouz2022safety} and oneshot time-varying~\cite{mazouz2026stochastic} baselines.

In short, the main contributions of this paper are:
\begin{itemize}
\item A sequential formulation of time-varying reach-avoid \emph{joint} controller-certificate synthesis for nonlinear discrete-time stochastic systems,
decomposed into a sequence of SDPs.
\item Two sequential reach-avoid certificate synthesis algorithms: an initial-set tracking formulation based on Monte Carlo propagation of finitely many samples from the initial set, and 
a uniform-sampling formulation that promotes maximizing the certificate throughout the safe set.

\item Benchmark comparisons with \emph{time-invariant} and \emph{oneshot} time-varying reach-avoid certificate formulations across systems of varying dimensions and time horizons, highlighting the trade-off between computational efficiency and certified reach-avoid probability guarantees.
\end{itemize}

\paragraph*{Related Work}
Certificate-based verification of stochastic systems has been studied extensively through \gls{sbfs}, which provide probabilistic safety guarantees over continuous state spaces via functional certificate conditions. Classical \gls{sbfs} formulations~\cite{santoyo2021barrier, jagtap2020formal, mazouz2022safety, yaghoubi2020risk, clark2021control, xue2025finite} employ a single time-invariant certificate across the entire horizon, yielding tractable convex optimization problems. However, these approaches often produce conservative  guarantees because a single function must certify safety over all time steps. To reduce this conservatism, piecewise certificate formulations partition the state space and synthesize multiple local certificates~\cite{mazouz2024piecewise, mazouz2025piecewise, mazouz2026stochasticbarrier}. While this increases the expressive power of the certificate, the formulation remains fundamentally time-invariant and therefore does not explicitly capture the temporal evolution of the stochastic dynamics. 

The reach-avoid problem generalizes safety by requiring the system to reach a target set while avoiding unsafe states throughout its evolution~\cite{summers2010verification}. Existing certificate-based approaches for stochastic reach-avoid verification over continuous state spaces remain relatively limited. These formulations are predominantly time-invariant and derive sufficient conditions for lower or upper bounds on reach-avoid probabilities~\cite{xue2025finite, vzikelic2023learning, wicker2024probabilistic}. Several works have also considered iterative controller--certificate synthesis using \gls{sos} optimization~\cite{mazouz2022safety, wajid2025successive, panja2025correct} or learning-based approaches~\cite{binny2026safe, salamati2021data, mazouz2024data, tayal2024learning}. These methods alternate between controller optimization and certificate synthesis and therefore generally converge to (local) sub-optimal solutions. More recently, a learning-based, data-driven approach was proposed for the joint synthesis of reach-avoid controllers and certificates~\cite{kong2026training}. However, that approach still uses a single certificate over the entire horizon. As a result, the approach does not exploit the inherent backward-recursive \gls{dp} structure.

Time-varying certificates produce substantially tighter probability bounds while preserving convex \gls{sos} synthesis~\cite{mazouz2026timevarying, mazouz2026stochasticbarrier}. However, existing approaches synthesize all time-indexed certificates jointly in a single \gls{sdp}, 
thereby limiting scalability. In this work, we address this challenge by introducing two surrogate objectives that enable the scalable synthesis of time-varying reach-avoid certificates through a sequence of fixed-size SOS programs while preserving the underlying backward-recursive \gls{dp} structure. To the best of our knowledge, this is the first certificate framework to develop time-varying reach-avoid certificates by explicitly exploiting this backward-recursive structure through sequential synthesis.

\section{Problem Formulation}

We consider a discrete-time, continuous-space system, governed by stochastic difference equation
\begin{equation}
\label{eq:system}
    \px_{k+1} = f(\px_k, u_k, \pw_k),
\end{equation}
where $\px_k \in X \subseteq \mathbb{R}^n$ denotes the system state and $u_k \in U \subset \mathbb{R}^m$ denotes the control input at time step $k \in \naturalszero$. The disturbance $\pw_k \in W \subseteq \mathbb{R}^w$ is an i.i.d. random variable with distribution $p_{\mathbf{w}}$, and is assumed to admit a closed-form moment generating function (e.g., $\pw_k \sim \mathcal{N}(0,\Sigma)$). The dynamics are described by the polynomial map $f: X \times U \times W \to \mathbb{R}^n$, which is continuous in all of its arguments.

The control input $u_k$ is determined by a time-dependent \emph{feedback controller} $\pi: X \times \naturals \to U$.
Given a state $x \in \mathbb{R}^n$ and a Borel-measurable set $X' \subseteq X$, the  transition kernel of System~\eqref{eq:system} is defined as
\begin{equation}
\label{eq:transition_kernel}
    T(X' \mid x,\pi) := \int_{\mathbb{R}^{\mathrm{w}}} \mathbf{1}_{X'}(f(x,\pi(x), \pw ) \, p_{\mathbf{w}}(w) \, dw,
\end{equation}
where $\mathbf{1}_{X'}(\cdot)$ is the indicator function with $\mathbf{1}_{X'}(x) = 1$ if $x \in X'$ and $0$ otherwise. From an initial state $x_0 \in \mathbb{R}^n$, kernel $T$ defines a unique probability measure $\mathrm{Pr}^{x_0}$ over trajectories of System~\eqref{eq:system}~\cite{bertsekas2004stochastic}, such that for  
any time step $k \in \mathbb{N}_0$, initial state $x_0 \in \mathbb{R}^n$, and measurable sets $X_0, X_k \subseteq X$, the measure satisfies
\begin{subequations}
\label{eq:prob measure}
\begin{align*}
    &\mathrm{Pr}^{x_0}[\pX_0 \in X_0] = \mathbf{1}_{X_0}(x_0), \\
    &\mathrm{Pr}^{x_0}[\pX_k \in X_k \mid \pX_{k-1} = x, \, \pi] = T(X_k \mid x, \pi(x)).
\end{align*}
\end{subequations}


We study temporal properties of System~\eqref{eq:system}, focusing specifically on \emph{reach-avoid} specifications. 

\begin{definition}[Probabilistic Reach-Avoid]
    \label{def: probabilistic reach-avoid}
    Let $X_\safe \subset X$ denote the safe set, $X_0 \subseteq X_\safe$ the initial set, and $X_\reach \subseteq X_\safe$ the target (reach) set. Define the unsafe set as $X_\unsafe := X \setminus X_\safe$. Given a feedback controller $\pi$ and horizon $\hor \in \mathbb{N}_0 \cup \{\infty\}$, the \emph{probabilistic reach-avoid} specification is defined as
    \begin{multline*}
        \Pfin = \inf_{x_0 \in X_0} \mathrm{Pr}^{x_0}[ (\exists k \leq \hor,\ \pX_k \in X_\reach) \\
        \wedge
        (\forall k' < k,\ \pX_{k'} \notin X_\unsafe)].
    \end{multline*}
    When $\hor = \infty$, we use the shorthand $\Pinf$.
\end{definition}

Reach-avoid properties are a fundamental building block for more expressive temporal specifications, including those formulated in temporal logics such as LTL and its variants. We consider the following verification and joint synthesis problem. 




\begin{problem}[Reach-Avoid Verification and Control Synthesis]
\label{prob}
    Consider System~\eqref{eq:system} with semi-algebraic control set $U$, and let $X_\safe \subset \reals^n$, $X_0 \subseteq X_\safe$, and $X_\reach \subseteq X_\safe$ be semi-algebraic sets defining the safe, initial, and target regions as in Definition~\ref{def: probabilistic reach-avoid}. 
    \begin{enumerate}
        \item[1.] (Verification) Given a feedback controller $\pi$, a horizon $\hor \in \mathbb{N}_0 \cup \{\infty\}$, and a probability threshold $\delta \in [0,1]$, verify whether
        \begin{align*}
            \Pfin \geq \delta.
        \end{align*}
    
        \item[2.] (Synthesis) 
        Given a horizon $\hor \in \mathbb{N}_0$ and a probability threshold $\delta \in [0,1]$, synthesize a controller $\pi^*$ such that 
        \begin{align*}
            \Pfinopt \geq \delta.
        \end{align*}
    \end{enumerate}
\end{problem}

Exact computation of $\Pfin$ is generally intractable for nonlinear stochastic systems with continuous state spaces. The above problems are challenging because the system is stochastic and evolves in a continuous state space under nonlinear dynamics $f$. Moreover, the reach-avoid objective requires reasoning about two distinct sets, $X_u$ and $X_r$ simultaneously, which further complicates the analysis. Existing approaches, including finite-state abstractions and certificate-based methods, address these challenges but remain limited by scalability and conservatism, respectively. In addition, \emph{jointly} synthesizing a feedback controller over a continuous control space $U \subset \reals^m$ and a certificate over state space $X$ further compounds the computational difficulty. These factors generally make the exact computation of reach-avoid probabilities intractable. 


To address these challenges, we synthesize time-varying reach-avoid certificates \emph{sequentially} through recursive optimization. Specifically, we introduce certificates defined as real-valued functions satisfying a set of sufficient conditions and synthesized recursively by exploiting the backward-recursive structure of the problem. We then show that such certificate yields a lower bound on $P_\textrm{r}(X_\reach, X_\unsafe,$ $X_0, H, \pi)$, thereby reformulating the verification problem as an optimization problem over a sequence of certificates that maximizes this bound. Moreover, we formulate the control synthesis as a joint optimization problem over both the certificates and the controller. By restricting the certificates to the class of polynomials, we show that both problems admit convex \gls{sos} relaxations that can be solved as a sequence of \gls{sdp}s.


\begin{remark}
The assumption of polynomial $f$ in ~\eqref{eq:system} is not restrictive, as general nonlinear dynamics can be reduced to polynomial dynamics through (piecewise) convex relaxations of $f$~\cite{mazouz2022safety}.
\end{remark}



\begin{remark}
While this paper focuses on reach-avoid properties, the proposed method extends straightforwardly to safety properties.
\end{remark}
\section{Preliminaries}

Our approach to generate certificates is grounded in \emph{time-varying} certificates and polynomial functional analysis.


\subsection{Time-Varying Certificates}
Time-varying reach-avoid certificates provide lower bounds on the probability of reaching a reach set while avoiding unsafe states. Unlike time-invariant certificates, which use a single function $\RA(x)$ for all time steps, time-varying certificates assign a distinct function $\RA(\cdot,i)$ to each step $i$, allowing a closer approximation of the value function. This reduces conservatism and yields tighter probability bounds, as shown in ~\cite{mazouz2026timevarying}. We recall the reach-avoid formulation.

\begin{theorem}[Time-Varying, Bounded-Horizon Reach-Avoid Certificate \cite{mazouz2026timevarying}]
\label{th:reach-avoid value function}
    Consider System~\eqref{eq:system} 
    under feedback controller $\pi$, 
    with a reach-avoid property defined by safe set $X_s$, initial set $X_0$, target set $X_\reach$, unsafe set $X_u$, and horizon $\hor \in \naturals_0$ per Definition~\ref{def: probabilistic reach-avoid}. Let $\nx = f(x, \pi, \textbf{w})$. Then, the real-valued function $\RA : \mathbb{R}^n \times \naturals^{\leq \hor}_{\geq 0} \to \mathbb{R}_{\geq 0}$ that is continuous in its first argument is a \emph{reach-avoid} certificate if, for some constants $\alpha_0, \alpha_i,\beta_i,\gamma \in \reals_{\geq 0}$ with $i \in \naturals^{\leq \hor}_{\geq 1}$, it satisfies 
    \begin{subequations}
        \label{eq: certificate tv finite H}
        \begin{align}
            \!\!\!\!\!&\RA(x, i) \geq 0 && \forall x \in \mathbb{R}^n, \ \forall i \in \naturals^{\leq \hor}_{\geq 0} \label{eq:nonneg-tv} \\
            &\RA(x, i) \leq 1 + \alpha_i && \forall x \in X_\reach, \ \forall i \in \naturals^{\leq \hor}_{\geq 0} \label{eq:reach-tv} \\
            &\RA(x, i) \leq \alpha_i && \forall x \in X_\unsafe, \ \forall i \in \naturals^{\leq \hor}_{\geq 0} \label{eq:unsafe-tv} \\
            &\RA(x, 0) \leq \alpha_0 && \forall x \in X_\safe\setminus X_\reach,  \label{eq:initial-nsafe-tv} \\
            &\RA(x, i) \leq \mathbb{E}[\RA(\px', i-1) | x ]  && \nonumber \\
            &  \qquad \qquad  - \alpha_{i-1} + \alpha_i + \beta_i 
                 && \forall x \in X_\safe \! \setminus \! X_\reach,  \forall i \in \naturals^{\leq \hor}_{\geq 1} \label{eq:exp-tv}\\
            &\RA(x, \hor) \geq \gamma && \forall x \in X_0. \label{eq:Vinitial_ss}
        \end{align}
    \end{subequations}
    Then, the reach-avoid probability is lower-bounded by
    \begin{align}
        \label{eq:tv-bound}
        \Pfin \geq \gamma - \left( \alpha_\hor + \sum_{i=1}^{ \hor} \beta_i
        \right ).
    \end{align}  
\end{theorem}


The time-varying certificates established in this theorem 
lower-bounds the true reach-avoid probability value.  Let $V_i(x)$ be the true probability of satisfying the reach-avoid requirement within $i$ steps starting from $x$.  Then, $\RA(x,i) - \alpha_i - \sum_{j=1}^i\beta_j \leq V_i(x)$.  


The conditions on $\RA$ are designed to facilitate certificate synthesis through functional optimization. In particular, the parameter $\alpha_i$ introduces a relaxation that allows $\RA(x,i)$ to be represented by a smooth function, such as a polynomial. To illustrate, consider $i=0$, for which $V_0(x)$ is necessarily an indicator function: $V_0(x)=1$ if $x\in X_\reach$ and $V_0(x)=0$ otherwise. Introducing $\alpha_0$ allows a smooth certificate $\RA(x,0)$ to lower-bound $V_0(x)$ up to the relaxation term, i.e.,
$\RA(x,0)-\alpha_0\leq V_0(x).$
This relaxation enables the use of polynomial certificates without requiring them to exactly capture the discontinuity of $V_0$.



\begin{remark}
Theorem~\ref{th:reach-avoid value function} readily extends to the infinite-horizon setting by imposing the additional martingale condition
$$\mathbb{E}[\RA(f(x,\pw), i-1)\mid x] \leq \RA(x, i)$$ at time index $i=H+1$. 
For further details, we refer the reader to~\cite{mazouz2026timevarying}.
\end{remark}


\subsection{Non-negative Polynomials}

To render the conditions into a tractable convex program, we work with polynomial certificates and enforce non-negativity on semi-algebraic sets via SOS relaxations.

\begin{definition}[SOS Polynomial]
\label{def:sos}
A multivariate polynomial $\lambda(x)$ is a SOS for $x \in \reals^n$ if there exist polynomials $\lambda_i$, $i = 1,\ldots,r$, for some $r \in \naturals$, such that
$\lambda(x) = \sum_{i=1}^{r} \lambda_i^2(x).$
If $\lambda$ is SOS, then $\lambda(x) \geq 0$ for all $x \in \reals^n$. The set of SOS polynomials is denoted by $\Lambda$.
\end{definition}

Verifying that a polynomial is SOS reduces to a semidefinite feasibility problem~\cite{parrilo2003semidefinite}. To enforce non-negativity on a compact basic semi-algebraic set rather than on all of $\reals^n$, we use the following standard result.

\begin{proposition}[\!{\cite{Stengle1974}}]
\label{prop:positivstellensatz}
Let $X = \{x \in \reals^n \mid h_i(x) \geq 0,\ i = 1,\ldots,l\}$ be a compact basic semi-algebraic set, where each $h_i$ is a polynomial. Then a polynomial $\gamma(x)$ is non-negative on $X$ if there exist $\lambda_0, \lambda_1, \ldots, \lambda_l \in \Lambda$ such that
$\gamma(x) = \lambda_0(x) + \sum_{i=1}^{l} \lambda_i(x)\, h_i(x).$
\end{proposition}

\begin{corollary}
\label{cor:sproc}
Let $h(x) = (h_1(x),\ldots,h_l(x))^\top$ describe $X$ as in Proposition~\ref{prop:positivstellensatz}, and let $\mathcal{L}(x) = (\lambda_1(x),\ldots,\lambda_l(x))^\top$ be a vector of SOS polynomials. Then $\gamma(x) - \mathcal{L}(x)^\top h(x) \in \Lambda$ implies $\gamma(x) \geq 0, \forall x \in X$.
\end{corollary}

Corollary~\ref{cor:sproc} provides the S-procedure encoding we use throughout the paper: each set-constrained inequality in Theorem~\ref{th:reach-avoid value function} is enforced by an SOS Lagrangian multiplier.

\section{Sequential Certificate Synthesis}

As presented in~\cite{mazouz2026timevarying}, 
the certificate in Theorem~\ref{th:reach-avoid value function} can be computed optimally using a \emph{oneshot} functional optimization problem.  
In that formulation, however, the computational complexity grows with the horizon, limiting scalability. In this section, we introduce the main contribution of this paper: a time-varying certificate framework that decomposes the centralized optimization problem into a sequence of smaller optimization problems. Specifically, we exploit the backward-recursive structure of the reach-avoid problem to formulate a sequence of fixed-size optimization problems that can be solved sequentially. This decomposition improves scalability, but it may also introduce sub-optimality.

%

We begin by presenting the centralized oneshot optimization problem,
which serves as the foundation for the subsequent sequential formulation, 
and 
in Section~\ref{sec: heuristics}, we introduce strategies to mitigate the sub-optimality issue.

\subsection{Oneshot Optimization}
\label{sec:oneshot}

The reach-avoid formulation in Theorem~\ref{th:reach-avoid value function} naturally leads to the following centralized  \emph{oneshot} functional optimization problem that maximizes the certified lower bound on the reach-avoid probability:
\begin{align}
    \label{eq: oneshopt optimization}
    \max_{\RA(\cdot,i),\gamma,{\alpha_i},{\beta_i}}
\quad \gamma-\left(\alpha_H+\sum_{i=1}^{H}\beta_i\right) \quad \text{s.t.} \quad \text{constraints \eqref{eq:nonneg-tv}-\eqref{eq:Vinitial_ss}.}
\end{align}
The oneshot formulation jointly optimizes all time-indexed certificates over the horizon, yielding the tightest probability bound within the admissible certificate class, especially in comparison to the time-invariant certificate.  However, the resulting problem is computationally challenging, as it involves optimizing multiple parameterized functions $\RA(x,i)$ for $i \in \naturals_{\ge 0}^{\leq H}$ simultaneously, subject to many constraints. 

For instance, when using an SOS optimization approach to solve the oneshot problem, the number of decision variables and constraints increases with the number of time-indexed certificates, leading to a massive computational costs. We analyze the runtime complexity  below.

\paragraph{Oneshot time-varying certificates SOS complexity.
}
The finite-horizon formulation introduces $(H{+}1)$ polynomials $\{\RA(x,i)\}_{i=0}^H$ and corresponding multipliers.
This yields $\approx 10 \times (H{+}1)$ SOS constraints in total,
resulting in an SDP with (roughly) $(H{+}1)$ blocks of size $b(d,n)=\binom{n+d}{d}$, where $b$ denotes the number of monomials of degree at most $d$ in $n$ variables, $d$ is the polynomial degree, and $n$ is the state-space dimension.
The overall worst-case complexity is 
$$\mathcal{O}\!\big((H{+}1)^3 b(d,n)^3\big).$$

The resulting computational complexity limits the scalability of the centralized formulation in Eq.~\eqref{eq: oneshopt optimization}, particularly for larger horizons. This motivates decomposing the centralized problem into a sequence of smaller optimization problems to improve scalability.

\subsection{Sequential Optimization}

A key observation is that the constraints in the centralized formulation in Eq.~\eqref{eq: oneshopt optimization} are largely separable with respect to the time steps.  All constraints except~\eqref{eq:exp-tv} impose conditions on the certificate $\RA(x,i)$ at a given time step $i$, independently of the certificates at the other time steps. The recursive expectation constraint~\eqref{eq:exp-tv}, however, introduces a dependency between consecutive certificates $\RA(x,i)$ and $\RA(x,i-1)$. Importantly, this dependency is unidirectional: once $\RA(x,i-1)$ is fixed, the constraint becomes independent of all other unknown certificates, allowing $\RA(x,i)$ to be synthesized through a sequence of smaller optimization problems.

This observation motivates a \emph{sequential synthesis framework}, in which the certificates are computed recursively, starting with $\RA(x,0)$ and proceeding forward in the certificate index until $\RA(x,H)$. Consider the setup in Theorem~\ref{th:reach-avoid value function}. Then, a \emph{sequential} time-varying real-valued function $\RA : \mathbb{R}^n \times \naturals^{\leq \hor}_{\geq 0} \to \mathbb{R}_{\geq 0}$ can be obtained by the following (three) consecutive and distinct optimization problems:
\begin{enumerate}
    \item[1.]    For $i=\{0,1\}$, jointly solve
    \begin{align}
    \label{eq:opt1}
    \min_{\RA(\cdot,0),\RA(\cdot,1),\alpha_0,\alpha_1,\beta_1} \phi_1(\alpha_0,\alpha_1,\beta_1) & \text{ \quad s.t. \quad  \eqref{eq:nonneg-tv}-\eqref{eq:exp-tv}}
    \end{align}

    \item[2.] For each $i \in \naturals^{< \hor}_{>1}$, separately solve 
    \begin{align}
    \label{eq:opt2}
    \min_{\RA(\cdot,i),\alpha_i,\beta_i} \phi_2(\alpha_i,\beta_i) & \text{ \quad s.t. \quad \eqref{eq:nonneg-tv}-\eqref{eq:unsafe-tv}, \eqref{eq:exp-tv}}
    \end{align}

    \item[3.] For $i=H$, solve
        \begin{align}
    \label{eq:opt3}
    \max_{\RA(\cdot,H),\alpha_H,\beta_H, \gamma} \phi_3(\alpha_H,\beta_H, \gamma) & \text{\quad s.t. \quad \eqref{eq:nonneg-tv}-\eqref{eq:unsafe-tv}, \eqref{eq:exp-tv}-\eqref{eq:Vinitial_ss}}
    \end{align}
\end{enumerate}
where $\phi_1,\phi_2,\phi_3$ are objective functions yet to be defined.

At each step $i>1$, the previously synthesized certificate is treated as fixed, and only the current certificate and its associated variables are optimized. Rather than jointly optimizing all certificates in a single centralized problem, the proposed decomposition therefore requires solving a sequence of smaller optimization problems. This substantially reduces the computational burden of each optimization step while preserving the original recursive structure.




A key \emph{challenge}, however, is that unlike the \emph{oneshot} formulation in Eq.~\eqref{eq: oneshopt optimization}, the \emph{sequential} approach does not directly optimize the final reach-avoid probability, since the certificates and slack variables at future time steps are not yet available. In particular, $\gamma = \sup_{x \in X_0} \RA(x,\hor)$ depends on the entire certificate sequence through the recursive expectation constraint. Ideally, each intermediate certificate would approximate the optimal value function over the entire state space, but optimizing such a global objective is computationally intractable. 

A straightforward alternative is to set the objective functions $\phi_1$ and $\phi_2$ in~\eqref{eq:opt1} and~\eqref{eq:opt2} to minimize the local slack variables $\beta_i$, i.e., $\phi_1 = \alpha_0 + \alpha_1 + \beta_1$ and $\phi_2 = \alpha_i + \beta_i$, and to maximize $\phi_3=\gamma-(\alpha_\hor+\beta_\hor)$ in~\eqref{eq:opt3}. Nevertheless, this greedy strategy is significantly suboptimal, as minimizing intermediate slack does not necessarily lead to an optimal value for the final reach-avoid probability bound. To address this limitation, we introduce two types of surrogate objective functions that improve the quality of sequentially synthesized certificates while preserving computational scalability.

\subsection{Surrogate Approaches}
\label{sec: heuristics}







As discussed above, a main challenge in sequential synthesis is that the terminal objective parameter $\gamma$ is only considered in the final step. Since $\gamma$ lower-bounds the terminal certificate $\RA(x,H)$ over the initial set $X_0$, intermediate optimization problems lack direct information about the ultimate objective. To address this limitation, we introduce two surrogate approaches that guide intermediate certificate synthesis while preserving the sequential structure.

\subsubsection{Initial-Set Tracking}
Recall that the sequential optimization problems in Eq.~\eqref{eq:opt1}-\eqref{eq:opt3} exploit the backward-recursive structure of the certificates. Specifically, the certificate $\RA(x,i)$ corresponds to the reach-avoid probability of System~\eqref{eq:system} at time step $k=H-i$. This observation motivates optimizing $\RA(x,i)$ over the set of states reachable from $X_0$ at time $H-i$, denoted by $X_{H-i}$. In other words, $X_{H-i}$ represents the propagation of $X_0$ under the system dynamics for $H-i$ time steps.
However, computing the exact propagation of $X_0$ under stochastic nonlinear dynamics is generally intractable.  Instead of exact computation, inspired by the success of the Monte Carlo tree search approaches \cite{coulom2006efficient}, we simply approximate $X_{H-i}$ using Monte Carlo simulation. 
The result is a finite set of samples $\tilde{X}_{H-i} = \{\tilde{x}_j\}_{j=1}^M$ that can be used as a surrogate for the initial set $X_{H-i}$ for the optimization of $\RA(x,i)$.
Introducing an auxiliary variable $\delta_i$ satisfying $\RA(\tilde{x}_j,i)\geq\delta_i$ for all $\tilde{x}_j \in \tilde{X}_{H-i} \cap (X_\safe \setminus X_\reach)$, we define the objective and additional constraints for the 
consecutive optimization problems:
\begin{enumerate}
    \item[1.]    
    For $i=\{0,1\}$, the objective function is
    \begin{align*}
        \max_{\RA(\cdot,0), \RA(\cdot,1), \alpha_0, \alpha_1,\beta_1, \delta_0, \delta_1} 
    \delta_0+\delta_1 -\alpha_0-\alpha_1-\beta_1
    \end{align*}
    The \emph{additional} constraints are: $$\RA(\tilde{x}_j,i)\geq \delta_i \quad  \forall \tilde{x}_j \in \tilde{X}_{H-i} \cap (X_\safe \setminus X_\reach).$$
    
    \item[2.] For each $i \in \naturals^{< \hor}_{>1}$, the objective function is
    \begin{align*}
    \max_{\RA(\cdot,i),\alpha_i,\beta_i, \delta_i} 
    \delta_i-\alpha_i-\beta_i
    \end{align*}
     The \emph{additional} constraints are:  $$\RA(\tilde{x}_j,i)\geq\delta_i \quad  \forall \tilde{x}_j \in \tilde{X}_{H-i} \cap (X_\safe \setminus X_\reach).$$

    \item[3.] For $i=H$, the objective remains
        \begin{align*}
    \max_{\RA(\cdot,H),\alpha_H,\beta_H, \gamma} 
    \gamma-\alpha_H-\beta_H
    \end{align*}
\end{enumerate}


This surrogate encourages larger certificate values along the sampled evolution of the initial set, thereby guiding intermediate synthesis toward improving the final objective.

\begin{remark}
Increasing the number of samples in $\tilde{X}_{H-i}$ improves approximation accuracy but introduces additional constraints, resulting in a trade-off between accuracy and computational complexity.
\end{remark}

\subsubsection{Uniform Spatial Optimization}
The second approach builds on the dynamic programming interpretation of $\RA(x,i)$ as a lower bound on the true reach-avoid probability, represented by the value function $V_i(x)$. From this perspective, and motivated by the Bellman principle of optimality, an ideal certificate should provide a tight lower bound on the value function throughout the safe set $X_\safe$, rather than only over the initial set $X_0$. This motivates optimizing the certificate over the entire safe set.
However, directly optimizing the certificate over $X_\safe$ is computationally challenging. Instead, we propose a uniform-sampling surrogate that promotes maximizing the certificate throughout $X_\safe$. Specifically, we uniformly sample $M$ points from $X_\safe$, yielding a finite set $\tilde{X}_\safe = \{\tilde{x}_j\}_{j=1}^N$. We introduce an auxiliary variable $\delta_i^j\geq 0$ satisfying $\RA(\tilde{x}_j,i)\geq\delta_i^j$. We define the objective and additional constraints for the 
consecutive optimization problems:

\begin{enumerate}
    \item[1.]    
    For $i=\{0,1\}$, the objective function is
    \begin{align*}
    \max_{\RA(\cdot,0), \RA(\cdot,1), \alpha_0, \alpha_1, \beta_1, \delta_0, \delta_1} 
   \sum_{j=1}^N(\delta_0^j + \delta_1^j) - N (\alpha_0 + \alpha_1 + \beta_1)
    \end{align*}
    The \emph{additional} constraints are:
$\RA(\tilde{x}_j,i)\geq\delta_i^j \quad \forall \tilde{x}_j\in\tilde{X}_\safe$.


    \item[2.] For each $i \in \naturals^{< \hor}_{>1}$, the objective function is
    \begin{align*}
    \max_{\RA(\cdot,i),\alpha_i,\beta_i, \delta_i} 
    \sum_{j=1}^N\delta_i^j-N(\alpha_i+\beta_i)
    \end{align*}
    The \emph{additional} constraints are: 
    $\RA(\tilde{x}_j,i)\geq\delta_i^j \quad \forall \tilde{x}_j\in\tilde{X}_\safe$.


    \item[3.] For $i=H$, the objective remains
        \begin{align*}
    \max_{\RA(\cdot,H),\alpha_H,\beta_H, \gamma} 
    (\gamma-\alpha_H-\beta_H)
    \end{align*}
\end{enumerate}





This surrogate promotes uniformly larger certificate values across the safe set, rather than focusing solely on states reachable from the initial set. The number of points $N$ controls the granularity of the surrogate objective and the associated computational cost.

\begin{remark}
If $\RA$ is sufficiently smooth, maximizing the polynomial over discrete sample points provides a good approximation to maximizing it throughout the safe set $X_\safe$. Increasing the sample count improves accuracy at the cost of additional constraints and variables.
\end{remark}


\begin{remark}
While the sequential approaches introduce suboptimality, the improved scalability allows for more expressive templates for certificate. As demonstrated in our experiments, combining surrogate objectives with expressive certificates yields tighter probability bounds than oneshot synthesis.
\end{remark}

\subsubsection{Correctness}

Both these surrogate objectives are applied at intermediate steps $i<\hor$, while the terminal step remains as the maximization of $\gamma-(\alpha_\hor+\beta_\hor)$. Importantly, these objectives modify only the optimization criterion, i.e., objective functions in~\eqref{eq:opt1}-\eqref{eq:opt2}, and preserve the certificate constraints. The following corollary establishes the correctness of both surrogate approaches in preserving the certified reach-avoid probability bound.

\begin{corollary}[Correctness of Surrogate Sequential Synthesis]
\label{cor:surrogate-correctness}
Consider System~\eqref{eq:system} and the reach-avoid certificate conditions~\eqref{eq: certificate tv finite H} of Theorem~\ref{th:reach-avoid value function}. Suppose the certificates $\RA(\cdot,i)$ are synthesized sequentially using either surrogate objective at each intermediate step $i<\hor$, while satisfying conditions~\eqref{eq:nonneg-tv}-\eqref{eq:exp-tv} at every step and~\eqref{eq:Vinitial_ss} at the terminal step. Then, the resulting certificate sequence satisfies
\begin{align*}
    \Pfin \geq \gamma-\left(\alpha_\hor+\sum_{i=1}^{\hor}\beta_i\right).
\end{align*}
\end{corollary}

\begin{proof}
Let $V_i(x)$ denote the reach-avoid value function over $i$ steps, satisfying the dynamic programming recursion
\begin{align*}
    V_i(x)=
    \begin{cases}
        1, & x\in X_\reach,\\
        0, & x\in X_\unsafe,\\
        0, & x\in X_\safe\setminus X_\reach,\ i=0,\\
        \mathbb{E}[V_{i-1}(\px')\mid x],
        & x\in X_\safe\setminus X_\reach,\ i>0.
    \end{cases}
\end{align*}
We show by induction that
$ V_i(x)\geq \RA(x,i)-\alpha_i-\sum_{j=1}^{i}\beta_j.
$.
For $i=0$, conditions~\eqref{eq:reach-tv}-\eqref{eq:initial-nsafe-tv} imply $V_0(x)\geq\RA(x,0)-\alpha_0$ for all $x$. Suppose the inequality holds at step $i-1$. For $x\in X_\safe\setminus X_\reach$, the dynamic programming recursion and~\eqref{eq:exp-tv} yield
\begin{align*}
    V_i(x)
    &\geq \mathbb{E}[\RA(\px',i-1)\mid x]
    -\alpha_{i-1}-\sum_{j=1}^{i-1}\beta_j\\
    &\geq \RA(x,i)-\alpha_i-\sum_{j=1}^{i}\beta_j.
\end{align*}
The boundary conditions~\eqref{eq:reach-tv} and~\eqref{eq:unsafe-tv} establish the same inequality on $X_\reach$ and $X_\unsafe$. Since the surrogate objectives do not alter these certificate conditions, the induction remains valid. Finally, condition~\eqref{eq:Vinitial_ss} gives
\begin{align*}
    \Pfin
    &\geq \inf_{x\in X_0}V_\hor(x)\\
    &\geq \inf_{x\in X_0}\RA(x,\hor)
    -\alpha_\hor-\sum_{i=1}^{\hor}\beta_i\\
    &\geq \gamma-\left(\alpha_\hor+\sum_{i=1}^{\hor}\beta_i\right).
\end{align*}
\end{proof}

In the next section, we derive \gls{sos} optimization formulations for the sequential synthesis framework, resulting in tractable surrogate semidefinite programs for sequential time-varying certificates.

\section{Sum-of-Squares Optimization Formulation}
\label{sec:sos}
We now derive the \gls{sos} formulations for sequential time-varying certificate and joint controller-certificate synthesis. Functional optimization via \gls{sos} enables the direct search for certificate functions over a polynomial function space.
Moreover, it enables the direct search for certificate functions through semidefinite programming \cite{santoyo2021barrier}. 
Accordingly, throughout this section, we restrict $\RA$ to the class of polynomial functions. We incorporate the two surrogate objectives introduced into the sequential optimization problems.

\subsection{Certificate Synthesis via SOS}

We first consider certificate synthesis for a given feedback controller $\pi$
(Problem~\ref{prob}.1). 
Specifically, we formulate the time-varying certificate conditions in Theorem~\ref{th:reach-avoid value function} as a sequence of \gls{sos} programs. 
At each time step, the preceding certificate is fixed, and the current certificate is optimized using either the initial-set tracking or uniform spatial surrogate objective. 
This yields a sequence of smaller optimization problems while preserving the certificate conditions.

\begin{theorem}[Sequential Time-Varying Reach-Avoid SOS Certificate]
\label{th:sequential-sos-certificate}
Consider polynomial functions $\{\RA(x,i)\}_{i=0}^{\hor}$ and semi-algebraic sets $X_j=\{x\in\reals^n\mid h_j(x)\geq0\}$, for $j\in\{\bar s,\initial,\unsafe,\reach\}$, where $X_{\bar\safe}=X_\safe\setminus X_\reach$ and each $h_j$ is a polynomial vector. Let $\Lambda$ denote the SOS cone, with $\mathcal{L}_{\bar\safe}^i$, $\mathcal{L}_\initial$, $\mathcal{L}_\unsafe^i$, and $\mathcal{L}_\reach^i$ denoting SOS multiplier vectors of appropriate dimensions. Given a polynomial controller $\pi$ and $\nx=f(x,\pi,\pw)$, a time-varying reach-avoid certificate can be synthesized sequentially by imposing the following SOS constraints:

\begin{subequations}
\label{eq:sequential-sos}
\begin{align}
&\RA(x,i)\in\Lambda, 
\label{constraint1-seq} \hspace{5cm}\\
&1+\alpha_i-\RA(x,i)
-\mathcal{L}_\reach^i(x)^\top h_\reach(x)\in\Lambda,
\label{constraint2-seq}\\
&\alpha_i-\RA(x,i)
-\mathcal{L}_\unsafe^i(x)^\top h_\unsafe(x)\in\Lambda,
\label{constraint3-seq}
\end{align}
\end{subequations}

\setcounter{equation}{\value{parentequation}}
\addtocounter{equation}{-1}
\begin{subequations}
\renewcommand{\theequation}{\theparentequation\alph{equation}}
\setcounter{equation}{4}
\begin{align}
&\alpha_0-\RA(x,0)
-\mathcal{L}_{\bar\safe}^0(x)^\top h_{\bar s}(x)\in\Lambda,
\label{constraint4-seq}
\\
&\mathbb{E}[\RA(\px',i-1)\mid x]
-\RA(x,i)-\alpha_{i-1}
+\alpha_i+\beta_i - \nonumber\\
&\hspace{4.5cm}
\mathcal{L}_{\bar\safe}^i(x)^\top h_{\bar s}(x)
\in\Lambda,
\label{constraint5-seq}\\
&\RA(x,\hor)-\gamma
-\mathcal{L}_\initial(x)^\top h_\initial(x)
\in\Lambda.
\label{constraint6-seq}
\end{align}
\end{subequations}

Here, constraints~\eqref{constraint1-seq}-\eqref{constraint3-seq} hold for $i\in \naturals^{\leq \hor}_{\geq 0}$, constraint~\eqref{constraint4-seq} is imposed at $i=0$, constraint~\eqref{constraint5-seq} holds for $i\in \naturals^{\leq \hor}_{\geq 1}$, and constraint~\eqref{constraint6-seq} is imposed at $i=\hor$.
The SOS constraints provide a semidefinite representations of the constraints~\eqref{eq:nonneg-tv}-\eqref{eq:Vinitial_ss}. With the objectives in~\eqref{eq:opt1}-\eqref{eq:opt3}, the resulting certificates guarantee
\begin{align}
    \Pfin\geq
    \gamma-\left(\alpha_\hor+\sum_{i=1}^{\hor}\beta_i\right).
\end{align}
\end{theorem}

\begin{proof}
By Corollary~\ref{cor:sproc}, constraints~\eqref{constraint1-seq}-\eqref{constraint6-seq} imply the corresponding certificate conditions~\eqref{eq:nonneg-tv}-\eqref{eq:Vinitial_ss} of Theorem~\ref{th:reach-avoid value function}. Since the surrogate objectives preserve these conditions, the resulting certificate sequence satisfies Theorem~\ref{th:reach-avoid value function}. 
\end{proof}

\paragraph{Sequential time-varying certificates SOS complexity.}
The sequential formulation decomposes the synthesis into $(\hor+1)$ smaller optimization problems, each involving a single certificate polynomial $\RA(x,i)$ and its corresponding multipliers. Each problem contains approximately $10$ SOS constraints, resulting in an SDP with blocks of size $b(d,n)$.
Since the problems are solved sequentially, the overall worst-case complexity is
\[
\mathcal{O}\!\big((\hor+1)b(d,n)^3\big).
\]

Compared to the oneshot complexity shown in Sec.~\ref{sec:oneshot}, sequential synthesis reduces the worst-case complexity from $\mathcal{O}((\hor+1)^3b(d,n)^3)$ to $\mathcal{O}((\hor+1)b(d,n)^3)$, yielding a quadratic reduction with respect to the horizon $\hor$. This improvement arises from solving smaller optimization problems sequentially rather than a single large SDP, substantially improving scalability for longer horizons.

Furthermore, the sequential formulation reduces the peak memory complexity from $\mathcal{O}((\hor+1)^2b(d,n)^2)$ to $\mathcal{O}(b(d,n)^2)$, yielding a quadratic reduction with respect to the horizon $\hor$. This follows from solving each SDP independently, requiring only the current optimization problem to be stored in memory.

\subsection{Joint Controller-Certificate SOS Synthesis}




We now extend the sequential certificate framework to the joint synthesis of a polynomial feedback controller and a time-varying reach-avoid certificate (Problem~\ref{prob}.2). Unlike verification, where the controller is given and only the certificate is optimized, synthesis requires optimizing both the controller and certificate functions jointly. Accordingly, we parameterize the controller $\pi_i$ as a polynomial function of the state $x$. 


In \cite{mazouz2026timevarying}, joint controller-certificate synthesis was formulated as a \emph{oneshot} optimization problem solved via dual SOS programming. However, the large number of decision variables limited its scalability. The proposed sequential formulation addresses this limitation by fixing the certificate from the preceding time step, allowing the controller to enter the recursive expectation constraint through a known polynomial. Similar to the verification problem, this enables joint synthesis through smaller SOS optimization problems with fewer decision variables, improving computational scalability and accommodating more expressive certificate templates. At intermediate steps, we employ the uniform spatial surrogate objective, which optimizes the certificate over the entire state space.

\begin{remark}
Initial-set tracking is not applicable to joint synthesis, as propagating the initial set (Monte Carlo simulation) requires knowledge of the controller, whose synthesis is part of the optimization problem. 
\end{remark}

\begin{figure*}[t]
    \centering
    \begin{minipage}{0.48\textwidth}
        \centering
        \includegraphics[width=\linewidth]{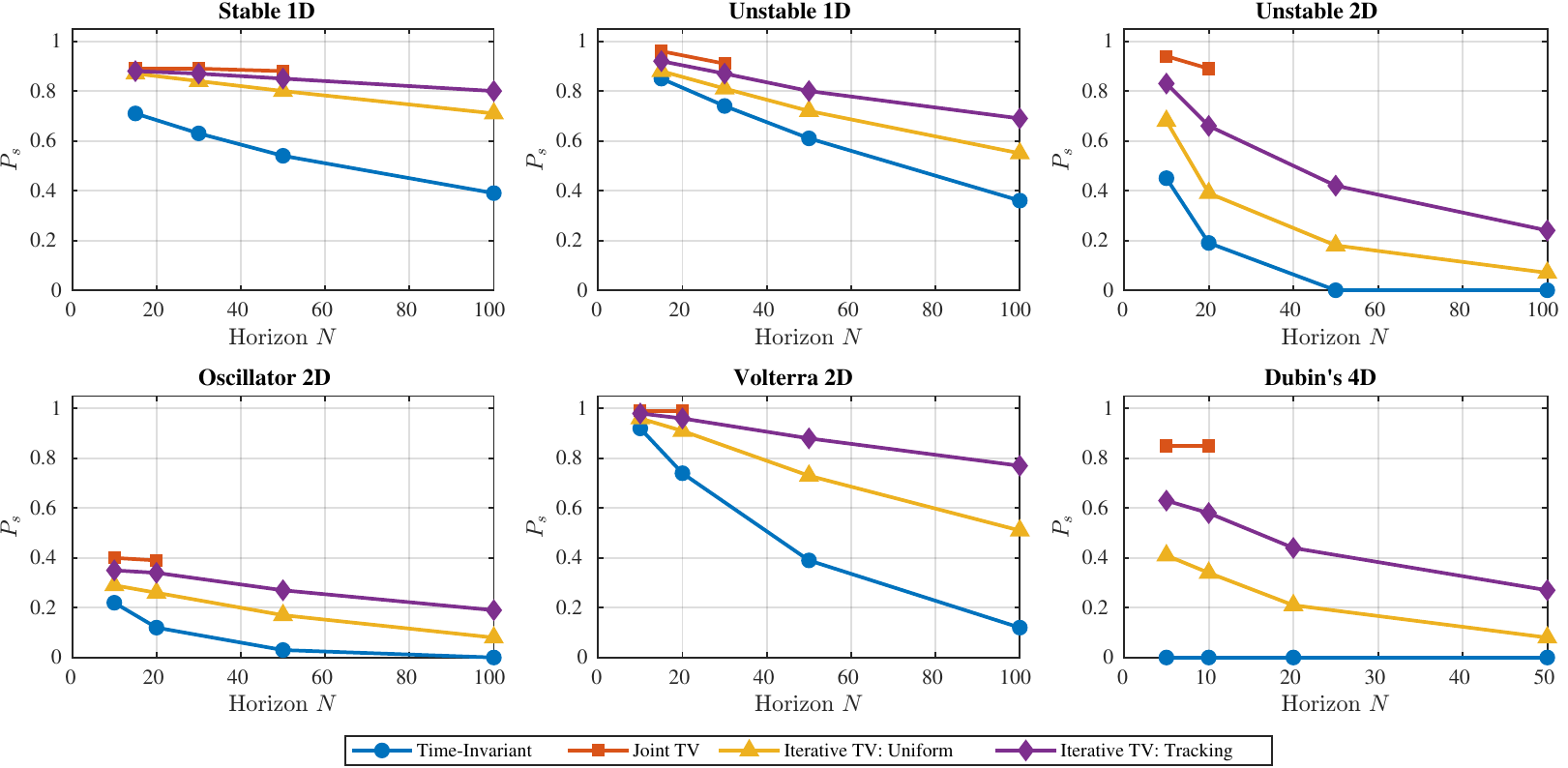}
        \caption{Certified lower bound $\underline{P}_\safe$ as a function of the horizon $\hor$ across benchmark systems. Missing markers correspond to \texttt{out-of-memory} or \texttt{time-out} failures.}
        \Description{}
        \label{fig:barrier_probability}
    \end{minipage}
    \hspace{0.02\textwidth}
    \begin{minipage}{0.48\textwidth}
        \centering
        \includegraphics[width=\linewidth]{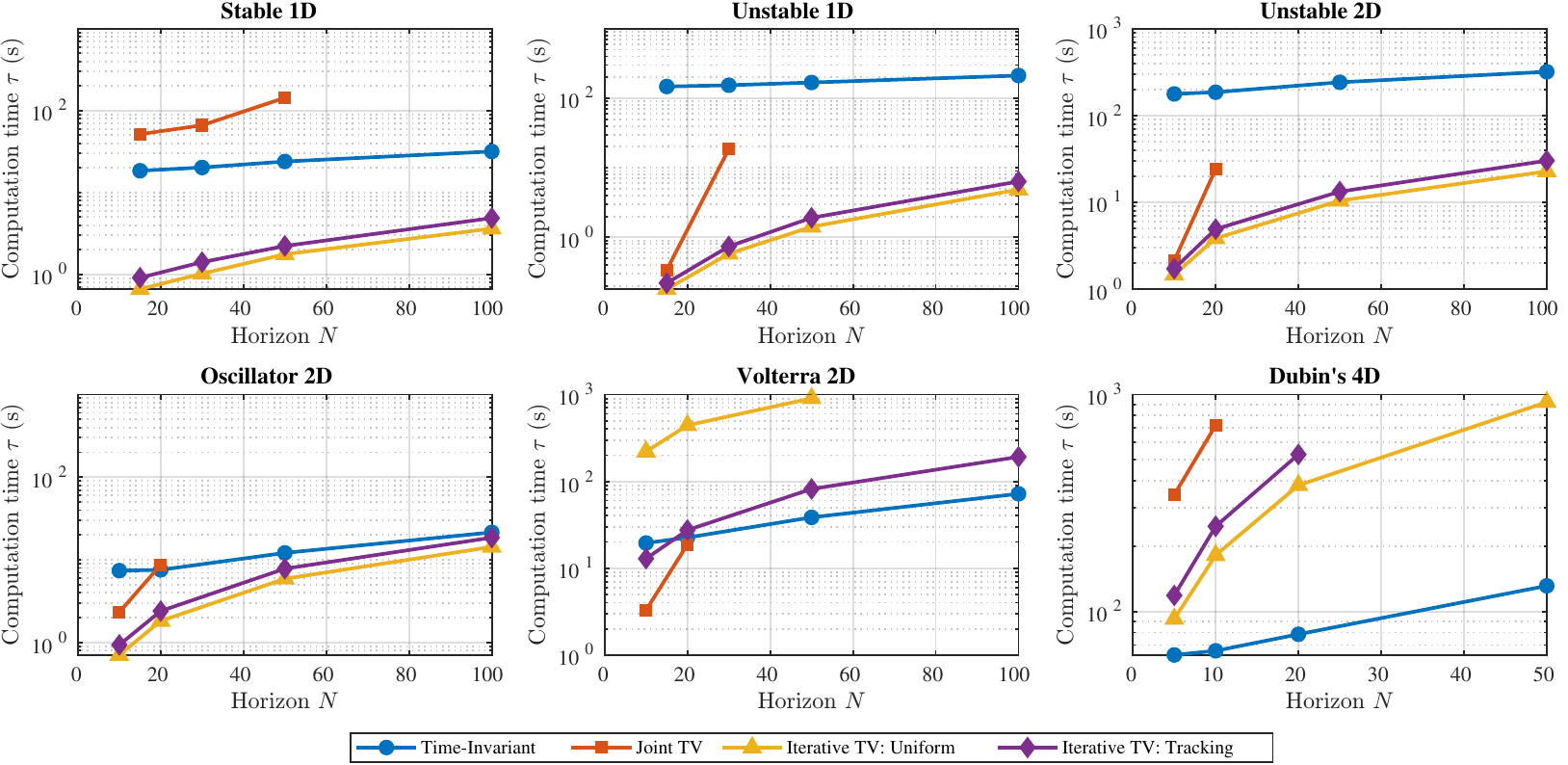}
        \caption{Computation time $\tau$ (logarithmic) as a function of the horizon $\hor$ across benchmark systems. The dashed horizontal line marks the 1000s time-out threshold.}
        \label{fig:barrier_computation_time}
        \Description{}
    \end{minipage}
\end{figure*}

We are now ready to formulate the joint controller-certificate optimization problem.

\begin{corollary}[Sequential Reach-Avoid Controller Synthesis]
\label{col:synthesis}
Consider System~\eqref{eq:system} with a reach-avoid property defined by safe set $X_\safe$, initial set $X_0$, target set $X_\reach$, unsafe set $X_\unsafe$, and horizon $\hor\in\naturals_0$. Let $\RA:\reals^n\times\naturals^{\leq\hor}_{\geq0}\to\reals_{\geq0}$ be continuous in its first argument.

A feedback controller sequence $\{\pi_i\}_{i=1}^{\hor}$ and its corresponding time-varying reach-avoid certificate can be synthesized by extending the sequential optimization problems in \eqref{eq:opt1}-\eqref{eq:opt3} to joint controller-certificate synthesis. Specifically, the initial certificate $\RA(x,0)$ and controller $\pi_1$ are jointly synthesized at the first step. For each subsequent intermediate step $i$, with $\RA(\cdot,i-1)$ fixed, we jointly optimize $\pi_i$, $\RA(\cdot,i)$, $\alpha_i$, and $\beta_i$ using a surrogate objective, subject to
\begin{subequations}
\label{eq:sequential-controller-synthesis}
\begin{align}
&\RA(x,i)\geq0,
&&\forall x\in\reals^n,\\
&\RA(x,i)\leq1+\alpha_i,
&&\forall x\in X_\reach,\\
&\RA(x,i)\leq\alpha_i,
&&\forall x\in X_\unsafe,\\
&\RA(x,i)\leq
\mathbb{E}\left[
\RA(f(x,\pi_i,\pw),i-1)\mid x
\right]\nonumber\\
&\qquad-\alpha_{i-1}+\alpha_i+\beta_i,
&&\forall x\in X_\safe\setminus X_\reach.
\end{align}
\end{subequations}
At the terminal step $i=\hor$, impose the same constraints together with
\begin{align}
    \RA(x,\hor)\geq\gamma,
    \qquad\forall x\in X_0,
\end{align}
and maximize $\gamma-(\alpha_\hor+\beta_\hor)$.

Any feasible controller-certificate sequence obtained through this procedure satisfies
\begin{align}
    \Pfin\geq
    \gamma-\left(
    \alpha_\hor+\sum_{i=1}^{\hor}\beta_i
    \right).
\end{align}
\end{corollary}

\begin{proof}
The proof follows from Theorem~\ref{th:reach-avoid value function}. At $i=0$, the synthesized certificate satisfies the required boundary conditions. At each subsequent step, the preceding certificate $\RA(\cdot,i-1)$ is fixed, and the jointly synthesized controller and current certificate satisfy the recursive expectation constraint and the corresponding boundary conditions. Thus, the resulting sequence satisfies all conditions of Theorem~\ref{th:reach-avoid value function}, regardless of the intermediate surrogate objectives. Imposing the terminal initial-set condition therefore yields
\begin{align}
    \Pfin\geq
    \gamma-\left(
    \alpha_\hor+\sum_{i=1}^{\hor}\beta_i
    \right).
\end{align}
\end{proof}


In what follows, we show that the sequential synthesis problem in Corollary~\ref{col:synthesis} admits a convex SOS formulation by leveraging the dual approach of~\cite{mazouz2026timevarying}. In particular, fixing the preceding certificate reduces our problem to a special case of the joint controller-certificate synthesis problem considered therein.

\begin{corollary}
Under the assumptions of Theorem 10 in~\cite{mazouz2026timevarying}, the SOS formulation of the sequential synthesis problem in Corollary~\ref{col:synthesis} admits a convex relaxation.
\end{corollary}
\begin{proof}
The result follows from Theorem 10 in~\cite{mazouz2026timevarying}. In particular, our sequential synthesis problem is a relaxed case of the joint controller-certificate synthesis problem considered therein, obtained by fixing the preceding certificate $\RA(\cdot,i-1)$. 
\end{proof}





\section{Results}
\label{sec:results}


In this section, we evaluate the proposed \emph{sequential} time-varying certificate synthesis procedures on a collection of stochastic benchmark systems ranging from one to six dimensions. Our experiments span safety and reach-avoid benchmarks, as well as ablation studies of key hyperparameters. We compare the performance metrics of four approaches: (i) time-invariant ~\cite{mazouz2022safety}, (ii) oneshot time-varying ~\cite{mazouz2026stochastic}, (iii) sequential time-varying SBFs with uniform subspace regularization, and (iv) sequential time-varying certificates with initial-set tracking. 
All methods are implemented in Julia using \texttt{SumOfSquares.jl}~\cite{weisser2019polynomial}. Experiments are performed on a machine with a 3.9 GHz 8-core processor and 128 GB of RAM. The description of the system setup can be found in App.~\ref{appendixA}.

\subsection{Safety Verification}

\begin{table*}[t!]
\caption{\emph{Highlighted} reach-avoid verification benchmark results for time-invariant certificates~\cite{mazouz2022safety}, one-shot time-varying certificates~\cite{mazouz2026stochastic}, and the proposed sequential time-varying certificates with uniform sampling and initial-set tracking. Here, $\underline{P}_\reach$ is the certified lower bound on the reach-avoid probability, $\hor$ the horizon, \texttt{deg} the certificate degree, and $\tau$ the computation time in seconds. \texttt{OOM} denotes an out-of-memory failure and \texttt{TO} denotes a 1000s time-out.}
\label{tab:reach-avoid-verification-highlight}
\centering
\resizebox{0.8\textwidth}{!}{
\begin{tabular}{l|c|ccc|ccc|ccc|ccc}
\toprule
\multirow{2}{*}{Model}
& \multirow{2}{*}{$\hor$}
& \multicolumn{3}{c|}{\texttt{Time-Invariant}}
& \multicolumn{3}{c|}{\texttt{One-Shot TV}~\cite{mazouz2026stochastic}}
& \multicolumn{3}{c|}{\texttt{Sequential TV: Uniform}}
& \multicolumn{3}{c}{\texttt{Sequential TV: Tracking}}
\\
&
& \texttt{deg} & $\underline{P}_\reach$ & $\tau(s)$
& \texttt{deg} & $\underline{P}_\reach$ & $\tau(s)$
& \texttt{deg} & $\underline{P}_\reach$ & $\tau(s)$
& \texttt{deg} & $\underline{P}_\reach$ & $\tau(s)$
\\
\midrule

Linear 1D
& 10
& 4 & 0.581 & 0.04
& 2 & 0.837 & 0.32
& 4 & 0.805 & 0.11
& 4 & 0.712 & 0.08 \\
& 10
& 12 & 0.795 & 0.17
& 10 & 0.996 & 9.58
& 20 & 0.997 & 3.89
& 20 & 0.995 & 2.94 \\
\cline{2-14}
& 20
& 6 & 0.686 & 0.07
& 4 & 0.952 & 1.49
& 8 & 0.965 & 0.59
& 8 & 0.906 & 0.44 \\
& 20
& 14 & 0.797 & 0.21
& 12 & 0.997 & 34.57
& 20 & 0.998 & 5.79
& 20 & 0.996 & 4.33 \\
\cline{2-14}
& 100
& 8 & 0.756 & 0.12
& 6 & 0.986 & 18.69
& 12 & 0.988 & 5.91
& 12 & 0.978 & 4.41 \\
& 100
& 14 & 0.823 & 0.32
& 12 & \textemdash & \texttt{OOM}
& 24 & 0.999 & 51.81
& 24 & 0.999 & 38.86 \\

\midrule

Contraction 2D
& 10
& 8 & 0.435 & 0.95
& 6 & 0.736 & 5.87
& 8 & 0.675 & 1.79
& 8 & 0.546 & 1.36 \\
& 10
& 28 & 0.786 & 23.81
& 24 & \textemdash & \texttt{TO}
& 32 & 0.995 & 148.83
& 32 & 0.986 & 111.63 \\
\cline{2-14}
& 20
& 12 & 0.567 & 2.89
& 10 & 0.826 & 41.34
& 12 & 0.846 & 6.87
& 12 & 0.776 & 5.22 \\
& 20
& 28 & 0.796 & 32.71
& 24 & \textemdash & \texttt{TO}
& 28 & 0.995 & 96.53
& 28 & 0.976 & 72.39 \\
\cline{2-14}
& 50
& 12 & 0.626 & 4.23
& 10 & 0.886 & 112.81
& 12 & 0.887 & 17.41
& 12 & 0.806 & 13.16 \\
& 50
& 28 & 0.816 & 48.71
& 24 & \textemdash & \texttt{TO}
& 28 & 0.995 & 243.82
& 28 & 0.986 & 182.54 \\

\midrule

Aircraft 3D
& 10
& 4 & 0.000 & 0.84
& 2 & 0.316 & 8.79
& 4 & 0.256 & 3.01
& 4 & 0.136 & 2.25 \\
& 10
& 12 & 0.066 & 18.51
& 10 & \textemdash & \texttt{OOM}
& 20 & 0.826 & 287.74
& 20 & 0.746 & 215.83 \\
\cline{2-14}
& 20
& 4 & 0.000 & 1.19
& 2 & 0.486 & 27.61
& 4 & 0.406 & 5.91
& 4 & 0.244 & 4.45 \\
& 20
& 12 & 0.086 & 26.93
& 10 & \textemdash & \texttt{OOM}
& 20 & 0.956 & 684.65
& 20 & 0.906 & 513.53 \\
\cline{2-14}
& 30
& 4 & 0.000 & 1.58
& 2 & 0.536 & 43.69
& 4 & 0.456 & 8.87
& 4 & 0.274 & 6.68 \\
& 30
& 10 & 0.056 & 14.91
& 8 & 0.906 & 864.38
& 16 & 0.946 & 492.74
& 16 & 0.886 & 369.58 \\

\midrule

Room Temp.\ 2D
& 10
& 4 & 0.126 & 4.89
& 2 & 0.516 & 51.94
& 4 & 0.467 & 11.49
& 4 & 0.367 & 8.77 \\
& 10
& 14 & 0.576 & 138.73
& 12 & \textemdash & \texttt{TO}
& 24 & 0.976 & 786.53
& 24 & 0.946 & 589.92 \\
\cline{2-14}
& 50
& 4 & 0.386 & 6.91
& 2 & 0.656 & 168.45
& 4 & 0.597 & 38.73
& 4 & 0.477 & 29.36 \\
& 50
& 10 & 0.724 & 56.83
& 8 & \textemdash & \texttt{OOM}
& 16 & 0.946 & 692.97
& 16 & 0.896 & 519.76 \\

\midrule

Room Temp.\ 3D
& 10
& 2 & 0.000 & 42.91
& 2 & 0.246 & 428.81
& 2 & 0.204 & 162.93
& 2 & 0.156 & 128.55 \\
& 10
& 6 & 0.000 & 326.95
& 6 & \textemdash & \texttt{TO}
& 6 & 0.716 & 926.94
& 6 & 0.636 & 914.73 \\

\bottomrule
\end{tabular}
}
\end{table*}


The safety verification results shown in Figs.~\ref{fig:barrier_probability} and~\ref{fig:barrier_computation_time} reveal a clear trade-off between the lower bound on the probability of safety and computational scalability, respectively. As expected, the oneshot time-varying formulation~\cite{mazouz2026stochastic} consistently achieves the tightest certified probability bound. By optimizing all certificates simultaneously, the joint formulation yields the least conservative guarantees. However,  
as the horizon increases, the size of the underlying SDP grows, leading to increases in runtime and, for several benchmark systems, out-of-memory failures. 

The time-invariant formulation exhibits the opposite behavior. The resulting bounds become increasingly conservative as the horizon grows, particularly for higher-dimensional systems. The certified safety probability decreases steadily with horizon length and often collapses to trivial values despite the use of high-degree barrier polynomials. This behavior highlights the limitations of time-invariant certificates in capturing the evolution of the reach-avoid probability over longer horizons, where a single certificate must satisfy the conditions across all time steps.

The sequential formulations provide an effective compromise. By synthesizing each stage independently, the size of the underlying \gls{sos} program remains fixed throughout the horizon, avoiding the horizon-dependent growth of the oneshot formulation while outperforming the time-invariant baseline. Initial-set tracking consistently yields stronger certificates than uniform sampling, particularly for linear systems with low-variance, zero-mean stochastic disturbances, where Monte Carlo trajectories remain concentrated around the nominal dynamics. Although sequential synthesis introduces some suboptimality through its surrogate objectives, the degradation in certified safety probabilities is modest relative to the computational gains, particularly for long horizons.

\subsection{Reach-Avoid Verification}

We next evaluate the sequential framework on the reach-avoid verification problem under higher noise variance, where the certificate must additionally account for the target set $X_\reach$. Table~\ref{tab:reach-avoid-verification-highlight} summarizes the certified probability bounds and computation times across different polynomial degrees and horizons. The full results are provided in Table~\ref{tab:reach-avoid-verification-full} of Appendix~\ref{app:full}.

The oneshot method reaches its computational limits earlier, timing out on the Contraction 2D system at $\hor=20$ and $\texttt{deg}=20$, and running out of memory on the Aircraft 3D system at $\texttt{deg}=10$. On Room Temp.\ 3D, it times out at degrees $4$ and $6$. The time-invariant certificate remains computationally inexpensive on the Contraction 2D and Aircraft 3D systems but produces substantially weaker bounds, certifying $\underline{P}_\reach=0$ for Aircraft 3D at degree $8$. In contrast, the sequential methods provide nontrivial bounds.

\begin{table*}[h!]
    \caption{Reach-avoid synthesis benchmark results for time-invariant certificates~\cite{mazouz2022safety}, one-shot time-varying certificates~\cite{mazouz2026stochastic}, and sequential time-varying certificates with uniform subspace regularization. Here, $\underline{P}_\reach$ denotes the reach-avoid probability lower bound, $\hor$ the horizon, \texttt{deg} the certificate degree, and $\tau$ the computation time in seconds. \texttt{OOM} denotes an out-of-memory failure and \texttt{TO} denotes a 1000s time-out for the reported benchmark runs. }
    \label{tab:reach-avoid-synthesis}
    \centering
    \resizebox{0.7\textwidth}{!}{
    \begin{tabular}{l|c|ccc|ccc|ccc}
        \toprule
        \multirow{2}{*}{Model}
        & \multirow{2}{*}{$\hor$}
        & \multicolumn{3}{c|}{\texttt{Time-Invariant}}
        & \multicolumn{3}{c|}{\texttt{One-Shot TV}~\cite{mazouz2026stochastic}}
        & \multicolumn{3}{c}{\texttt{Sequential TV: Uniform}}
        \\
        &
        & \texttt{deg} & $\underline{P}_\reach$ & $\tau(s)$
        & \texttt{deg} & $\underline{P}_\reach$ & $\tau(s)$
        & \texttt{deg} & $\underline{P}_\reach$ & $\tau(s)$
        \\
\midrule

Contraction 2D
& 10
& 6 & 0.681 & 38.21
& 4 & 0.894 & 71.36
& 6 & 0.739 & 17.82 \\
& 10
& 8 & 0.764 & 12.68
& 6 & 0.937 & 119.43
& 10 & 0.913 & 55.74 \\
& 10
& 10 & 0.821 & 19.94
& 8 & 0.973 & 214.61
& 14 & 0.958 & 174.28 \\
& 10
& 14 & 0.876 & 29.83
& 10 & 0.988 & 427.13
& 18 & 0.993 & 361.72 \\
\cline{2-11}
& 20
& 8 & 0.742 & 15.89
& 6 & 0.954 & 1318.46
& 10 & 0.938 & 121.37 \\
& 20
& 10 & 0.807 & 22.74
& 8 & 0.985 & 2917.63
& 14 & 0.978 & 349.85 \\
& 20
& 14 & 0.861 & 36.12
& 10 & \textemdash & \texttt{OOM}
& 18 & 0.995 & 702.18 \\
\cline{2-11}
& 50
& 8 & 0.826 & 23.16
& 6 & 0.976 & 3349.27
& 10 & 0.961 & 378.52 \\
& 50
& 10 & 0.879 & 30.12
& 8 & \textemdash & \texttt{OOM}
& 14 & 0.987 & 891.43 \\
& 50
& 14 & 0.934 & 44.08
& 10 & \textemdash & \texttt{OOM}
& 18 & 0.998 & 1893.26 \\

\midrule

Aircraft 3D
& 10
& 4 & 0.091 & 50.82
& 2 & 0.912 & 486.31
& 4 & 0.589 & 153.72 \\
& 10
& 6 & 0.137 & 130.24
& 4 & 0.948 & 1726.58
& 8 & 0.886 & 469.13 \\
& 10
& 8 & 0.194 & 293.47
& 6 & \textemdash & \texttt{OOM}
& 10 & 0.941 & 806.52 \\
\cline{2-11}
& 20
& 4 & 0.134 & 74.96
& 2 & 0.953 & 1014.28
& 4 & 0.706 & 235.84 \\
& 20
& 6 & 0.191 & 190.38
& 4 & 0.976 & 3368.42
& 8 & 0.961 & 733.91 \\
& 20
& 8 & 0.257 & 423.61
& 6 & \textemdash & \texttt{OOM}
& 10 & 0.985 & 1507.36 \\
\cline{2-11}
& 30
& 4 & 0.176 & 99.37
& 2 & 0.966 & 1523.84
& 4 & 0.769 & 311.46 \\
& 30
& 6 & 0.231 & 252.73
& 4 & 0.983 & 4397.21
& 8 & 0.981 & 958.62 \\
& 30
& 8 & 0.298 & 562.19
& 6 & \textemdash & \texttt{OOM}
& 10 & 0.994 & 2036.74 \\

\midrule

Room Temp.\ 3D
& 10
& 4 & 0.094 & 66.73
& 2 & 0.461 & 524.38
& 4 & 0.529 & 222.84 \\
& 10
& 6 & 0.131 & 153.26
& 4 & 0.637 & 1912.76
& 6 & 0.684 & 475.19 \\
& 10
& 8 & 0.196 & 335.82
& 6 & \textemdash & \texttt{OOM}
& 8 & 0.791 & 948.63 \\

\bottomrule
    \end{tabular}
    }
\end{table*}


Among the sequential variants, uniform subspace regularization consistently provides tighter certificates than initial-set tracking, at the cost of additional computation. Increasing the polynomial degree allows the sequential methods to match or exceed the bounds obtained by the oneshot formulation at lower degrees, while often requiring less computation time. For example, for Aircraft 3D with $\hor=20$, uniform regularization achieves $\underline{P}_\reach=0.746$ at degree $8$ in $22.63$s, compared with $0.826$ at degree $6$ in $263.48$s for the oneshot formulation. Increasing the sequential degree to $12$ further improves the bound to $0.956$ in $684.65$s.

Increasing the certificate degree generally improves the certified bound at a fixed horizon, while increasing the horizon yields tighter bounds at comparable degrees. The sequential formulation also accommodates higher degrees after the oneshot formulation exceeds the computational budget. For example, on Contraction 2D with $\hor=50$, uniform regularization achieves a bound of $0.995$ at degree $28$ in $243.82$s, whereas the oneshot formulation times out at degree $24$. On Room Temp.\ 3D, uniform regularization reaches $0.716$ at degree $6$ in $926.94$s, compared with $0.636$ in $914.73$s for tracking. Overall, the sequential approach compensates for the conservatism introduced by stage-wise optimization through higher-degree certificates, extending the feasible frontier.



\subsection{Reach-Avoid Synthesis}

Next, we consider the reach-avoid control synthesis problem, in which a feedback controller and the corresponding certificate are synthesized jointly. All methods use a polynomial controller of degree $2$. The sequential method synthesizes the controller stage-wise: at each stage $i$, with $\RA(\cdot,i-1)$ fixed, a parameterized polynomial controller $\pi_i$ is optimized together with $\RA(\cdot,i)$. In contrast, the oneshot and time-invariant baselines optimize a single coupled program over all controller and certificate parameters simultaneously. Table~\ref{tab:reach-avoid-synthesis} shows the certified probability bounds and computation times across different polynomial degrees.

Joint synthesis introduces additional computational overhead relative to verification, since the controller parameters add decision variables. This overhead is particularly pronounced in the oneshot formulation. For Contraction 2D at $\hor=20$, the oneshot method achieves $\underline{P}_\reach=0.985$ at degree $8$ in $2917.63$s, whereas sequential synthesis achieves $\underline{P}_\reach=0.978$ at degree $14$ in $349.85$s. Increasing the sequential degree to $18$ further improves the bound to $0.995$ in $702.18$s, while the oneshot formulation runs out of memory at degree $10$. The computational advantage becomes more pronounced on Aircraft 3D, where sequential synthesis achieves $\underline{P}_\reach=0.961$ at $\hor=20$ and degree $8$ in $733.91$s, compared with $0.976$ at degree $4$ in $3368.42$s for the oneshot formulation.

Table~\ref{tab:reach-avoid-synthesis} further illustrates how increasing the polynomial degree allows sequential synthesis to recover or exceed the bounds obtained by oneshot optimization at lower degrees, at the expense of additional computation. For Room Temp.\ 3D at $\hor=10$, the sequential bound increases from $0.529$ at degree $4$ to $0.791$ at degree $8$, while runtime increases from $222.84$s to $948.63$s. At longer horizons, sequential synthesis remains tractable where the baseline formulations exceed their computational budgets. For Aircraft 3D, increasing the horizon from $10$ to $30$ improves the sequential bound from $0.941$ at degree $10$ to $0.994$ at the same degree. Overall, the sequential formulation compensates for the suboptimality introduced by stage-wise synthesis through higher-degree certificates, extending the feasible frontier of reach-avoid control synthesis.


\subsection{Scalability}

\begin{figure}[t]
    \centering
    \includegraphics[width=\columnwidth]{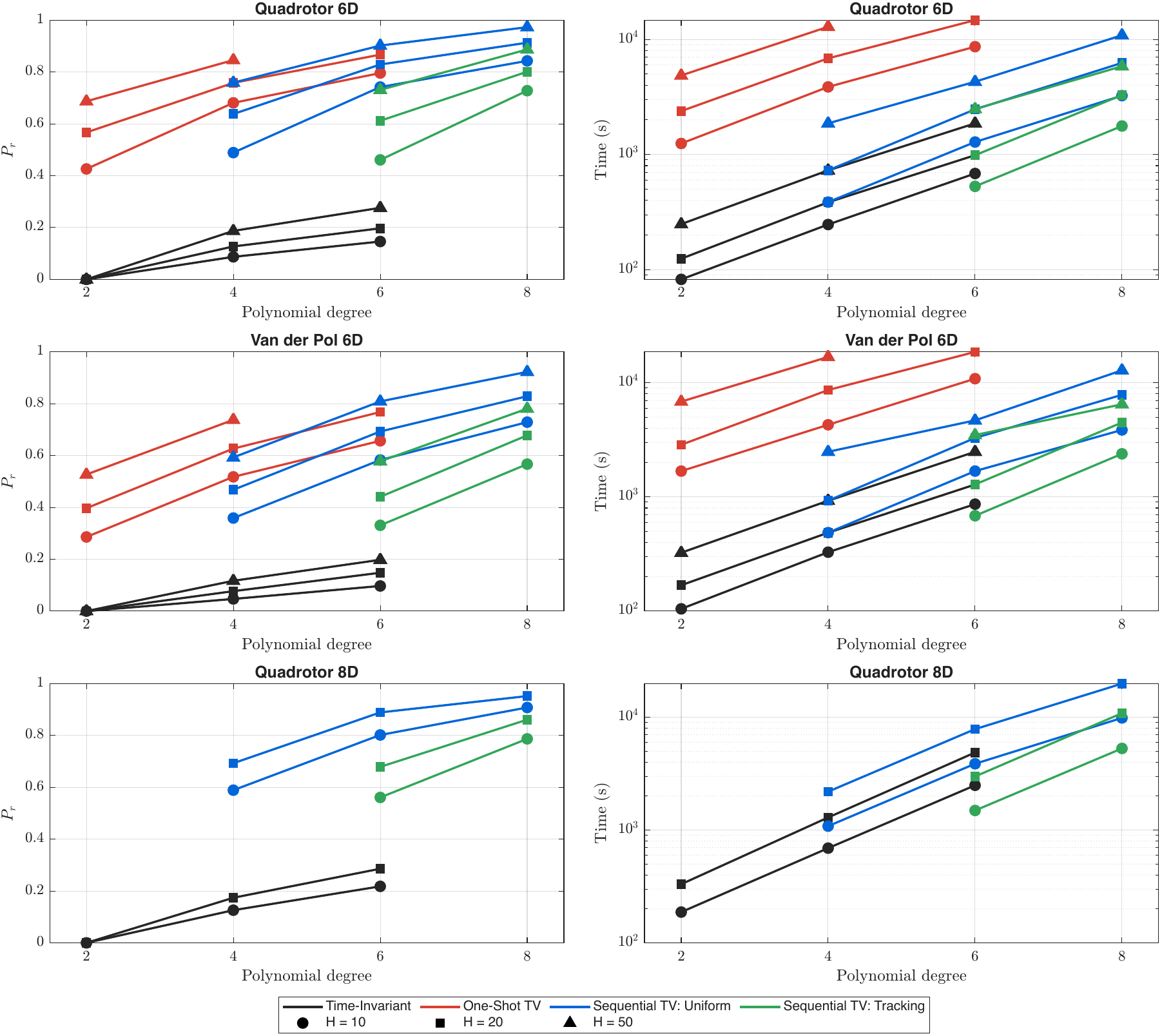}
    \caption{\textit{Verification.} Certified lower bound $\underline{P}_r$ as a function of the polynomial degree across benchmark systems. Missing markers correspond to \texttt{out-of-memory} failures.}
    \Description{}
    \label{fig:reach_avoid_verification}
\end{figure}

\begin{figure}[t]
    \centering
    \includegraphics[width=\columnwidth]{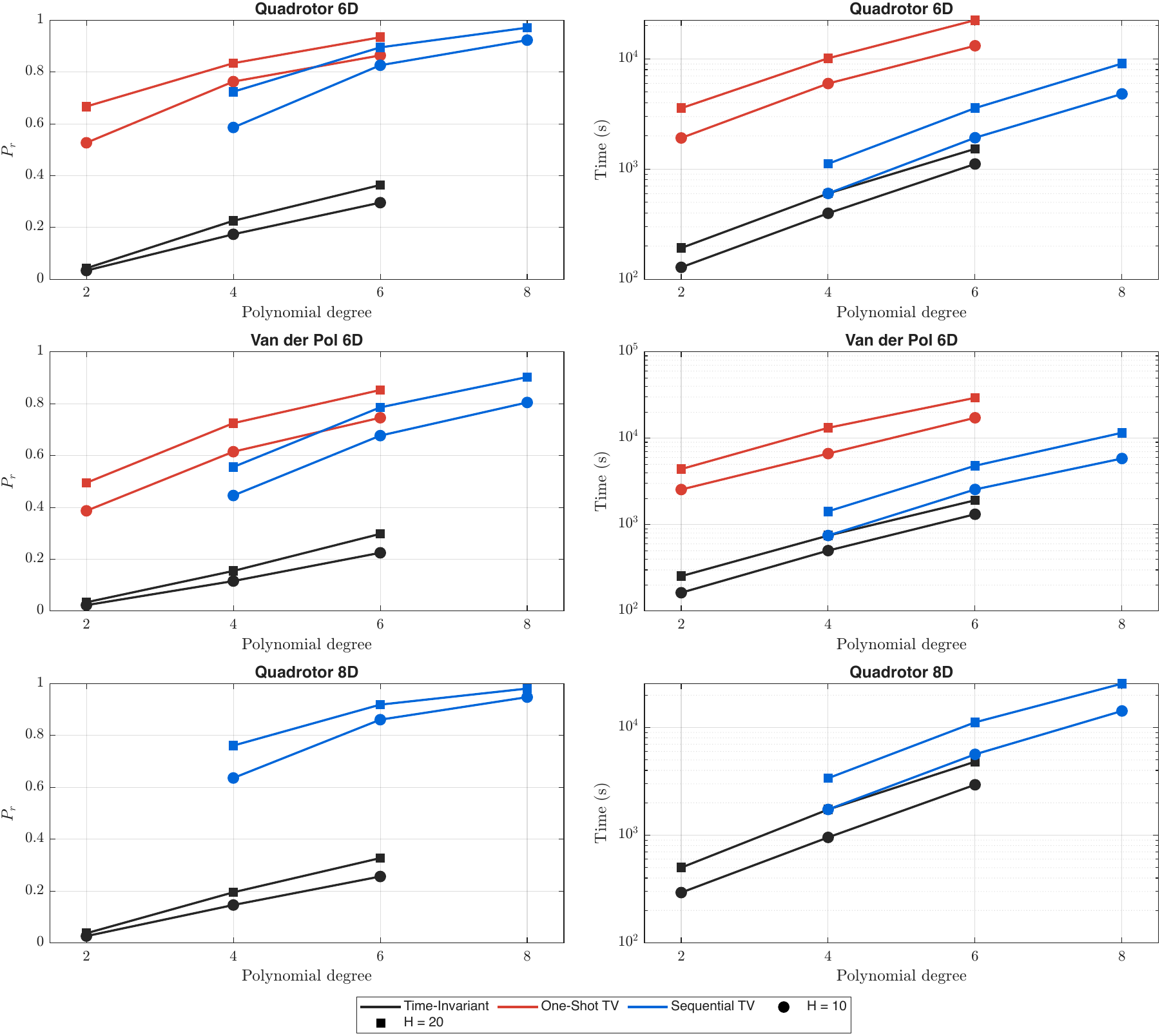}
    \caption{\textit{Synthesis.} Certified lower bound $\underline{P}_r$ as a function of the polynomial degree across benchmark systems. Missing markers correspond to \texttt{out-of-memory} failures.}
    \Description{}
    \label{fig:reach_avoid_synthesis}
\end{figure}

Finally, to further assess the scalability of the sequential framework, we consider higher-dimensional systems, including the Quadrotor and Van der Pol systems in six dimensions and the Quadrotor system in eight dimensions. Figs.~\ref{fig:reach_avoid_verification} and~\ref{fig:reach_avoid_synthesis} summarize the verification and synthesis results, respectively, in terms of certified reach-avoid probability bounds as a function of polynomial degree.

For verification, the oneshot formulation becomes increasingly computationally demanding as the system dimension and polynomial degree increase. In particular, it runs out of memory for all tested degrees on Quadrotor 8D and encounters memory limitations on the six-dimensional systems at higher degrees and horizons. In contrast, sequential synthesis successfully computes certificates for these problems. For instance, on Quadrotor 8D with $\hor=20$, uniform sampling achieves a certified bound of $0.952$ at degree $8$, whereas the oneshot formulation runs out of memory even at degree $2$. Similar trends are observed across other systems.

The synthesis results exhibit similar trends. The oneshot formulation runs out of memory even at the lowest tested degree on Quadrotor 8D, whereas sequential synthesis successfully computes joint controller-certificate pairs across all tested degrees and horizons, achieving a certified bound of $0.978$ at degree $8$ for $\hor=20$. It also provides tighter bounds at lower computational cost on the six-dimensional systems. Overall, these results demonstrate that sequential optimization substantially improves scalability for both verification and synthesis, enabling higher-dimensional problems and more expressive certificate templates that are computationally intractable with oneshot synthesis.

\section{Conclusion}

This paper develops a scalable framework for joint controller and certificate synthesis for discrete-time stochastic systems. Building on the dynamic programming characterization of time-varying certificates, we introduce a sequential formulation that decomposes the horizon-wide synthesis problem into a sequence of fixed-size sum-of-squares programs. To mitigate the suboptimality introduced by sequential decomposition, we propose two surrogate approaches: uniform sampling over the safe set and initial-set tracking through Monte Carlo propagation of finitely many samples. The framework preserves formal probabilistic guarantees while improving computational scalability and accommodating more expressive certificate templates. Numerical experiments on linear and nonlinear benchmark systems demonstrate substantially better scalability than joint oneshot synthesis and consistently tighter probability bounds than time-invariant certificates, with competitive or improved reach-avoid probability bounds relative to the oneshot formulation.


\bibliographystyle{acm}
\bibliography{cite}

\begin{thebibliography}{10}

\bibitem{abate2020arch}
{\sc Abate, A., Blom, H.~A., Cauchi, N., Delicaris, J., Hartmanns, A., Khaled, M., Lavaei, A., Pilch, C., Remke, A., Schupp, S., et~al.}
\newblock Arch-comp20 category report: Stochastic models.
\newblock In {\em ARCH\/} (2020).

\bibitem{bertsekas2004stochastic}
{\sc Bertsekas, D.~P., and Shreve, S.~E.}
\newblock {\em {Stochastic Optimal Control: the Discrete-time Case}}, vol.~5.
\newblock Athena Scientific, 2004.

\bibitem{binny2026safe}
{\sc Binny, A.~E., Anand, M., Kussaba, H.~T., Chen, L., Agrawal, S., Abu-Dakka, F.~J., and Swikir, A.}
\newblock Safe and stable neural network dynamical systems for robot motion planning.
\newblock {\em IEEE Robotics and Automation Letters\/} (2026).

\bibitem{clark2021control}
{\sc Clark, A.}
\newblock Control barrier functions for stochastic systems.
\newblock {\em Automatica 130\/} (2021), 109688.

\bibitem{coulom2006efficient}
{\sc Coulom, R.}
\newblock Efficient selectivity and backup operators in monte-carlo tree search.
\newblock In {\em International conference on computers and games\/} (2006), Springer, pp.~72--83.

\bibitem{guiochet2017safety}
{\sc Guiochet, J., Machin, M., and Waeselynck, H.}
\newblock Safety-critical advanced robots: A survey.
\newblock {\em Robotics and Autonomous Systems 94\/} (2017).

\bibitem{jagtap2020formal}
{\sc Jagtap, P., Soudjani, S., and Zamani, M.}
\newblock Formal synthesis of stochastic systems via control barrier certificates.
\newblock {\em IEEE Transactions on Automatic Control 66}, 7 (2020), 3097--3110.

\bibitem{kong2026training}
{\sc Kong, C.-W., Escobar, S., Gracia, I., McMahon, J., and Lahijanian, M.}
\newblock Training with hard constraints: Learning neural certificates and controllers for sdes.
\newblock {\em arXiv preprint arXiv:2602.23526\/} (2026).

\bibitem{kushner1967stochastic}
{\sc Kushner, H.~J.}
\newblock {Stochastic Stability and Control}.
\newblock Tech. rep., Brown Univ Providence RI, 1967.

\bibitem{legat2017sos}
{\sc Legat, B., Coey, C., Deits, R., Huchette, J., and Perry, A.}
\newblock {Sum-of-squares optimization in Julia}.
\newblock In {\em The First Annual JuMP-dev Workshop\/} (2017).

\bibitem{mazouz2025piecewise}
{\sc Mazouz, R., Laurenti, L., and Lahijanian, M.}
\newblock Piecewise control barrier functions for stochastic systems.
\newblock In {\em 2025 IEEE 64th Conference on Decision and Control (CDC)\/} (2025), IEEE, pp.~3571--3576.

\bibitem{mazouz2026stochastic}
{\sc Mazouz, R., Laurenti, L., and Lahijanian, M.}
\newblock Stochastic barrier certificates in the presence of dynamic obstacles.
\newblock {\em arXiv preprint arXiv:2604.20208\/} (2026).

\bibitem{mazouz2026timevarying}
{\sc Mazouz, R., Laurenti, L., and Lahijanian, M.}
\newblock Time-varying reach-avoid control certificates for stochastic systems.
\newblock {\em arXiv preprint arXiv:2603.25816\/} (2026).

\bibitem{mazouz2024piecewise}
{\sc Mazouz, R., Mathiesen, F.~B., Laurenti, L., and Lahijanian, M.}
\newblock Piecewise stochastic barrier functions.
\newblock {\em Automatica 189\/} (2026), 113020.

\bibitem{mazouz2026stochasticbarrier}
{\sc Mazouz, R., Mathiesen, F.~B., Laurenti, L., and Lahijanian, M.}
\newblock Stochasticbarrier. jl: A toolbox for stochastic barrier function synthesis.
\newblock {\em arXiv preprint arXiv:2602.20359\/} (2026).

\bibitem{mazouz2022safety}
{\sc Mazouz, R., Muvvala, K., Ratheesh~Babu, A., Laurenti, L., and Lahijanian, M.}
\newblock Safety guarantees for neural network dynamic systems via stochastic barrier functions.
\newblock {\em Advances in Neural Information Processing Systems 35\/} (2022), 9672--9686.

\bibitem{mazouz2024data}
{\sc Mazouz, R., Skovbekk, J., Mathiesen, F.~B., Frew, E., Laurenti, L., and Lahijanian, M.}
\newblock Data-driven permissible safe control with barrier certificates.
\newblock In {\em 2024 IEEE 63rd Conference on Decision and Control (CDC)\/} (2024), IEEE, pp.~6844--6849.

\bibitem{oumer2024safety}
{\sc Oumer, M.~A., Murali, V., Trivedi, A., and Zamani, M.}
\newblock Safety verification of discrete-time systems via interpolation-inspired barrier certificates.
\newblock {\em IEEE Control Systems Letters 8\/} (2024), 3183--3188.

\bibitem{panja2025correct}
{\sc Panja, P., and Platzer, A.}
\newblock Correct-by-construction barrier certificate synthesis for safety verification of continuous dynamical systems.
\newblock In {\em 2025 IEEE 64th Conference on Decision and Control (CDC)\/} (2025), IEEE, pp.~7701--7707.

\bibitem{parrilo2003semidefinite}
{\sc Parrilo, P.~A.}
\newblock Semidefinite programming relaxations for semialgebraic problems.
\newblock {\em Mathematical programming 96}, 2 (2003), 293--320.

\bibitem{salamati2021data}
{\sc Salamati, A., Lavaei, A., Soudjani, S., and Zamani, M.}
\newblock Data-driven verification and synthesis of stochastic systems through barrier certificates.
\newblock {\em arXiv preprint arXiv:2111.10330\/} (2021).

\bibitem{santoyo2021barrier}
{\sc Santoyo, C., Dutreix, M., and Coogan, S.}
\newblock A barrier function approach to finite-time stochastic system verification and control.
\newblock {\em Automatica 125\/} (2021), 109439.

\bibitem{Stengle1974}
{\sc Stengle, G.}
\newblock A nullstellensatz and a positivstellensatz in semialgebraic geometry.
\newblock {\em Mathematische Annalen 207}, 2 (1974), 87--97.

\bibitem{summers2010verification}
{\sc Summers, S., and Lygeros, J.}
\newblock Verification of discrete time stochastic hybrid systems: A stochastic reach-avoid decision problem.
\newblock {\em Automatica 46}, 12 (2010), 1951--1961.

\bibitem{tayal2024learning}
{\sc Tayal, M., Zhang, H., Jagtap, P., Clark, A., and Kolathaya, S.}
\newblock Learning a formally verified control barrier function in stochastic environment.
\newblock In {\em 2024 IEEE 63rd Conference on Decision and Control (CDC)\/} (2024), IEEE, pp.~4098--4104.

\bibitem{wajid2025successive}
{\sc Wajid, R., and Sankaranarayanan, S.}
\newblock Successive control barrier functions for nonlinear systems.
\newblock In {\em Proceedings of the 28th ACM International Conference on Hybrid Systems: Comp. and Control\/} (2025).

\bibitem{weisser2019polynomial}
{\sc Weisser, T., Legat, B., Coey, C., Kapelevich, L., and Vielma, J.~P.}
\newblock Polynomial and moment optimization in julia and jump.
\newblock In {\em JuliaCon}. 2019.

\bibitem{wicker2024probabilistic}
{\sc Wicker, M., Laurenti, L., Patane, A., Paoletti, N., Abate, A., and Kwiatkowska, M.}
\newblock Probabilistic reach-avoid for bayesian neural networks.
\newblock {\em Artificial Intelligence 334\/} (2024), 104132.

\bibitem{xue2025finite}
{\sc Xue, B.}
\newblock Finite-time safety and reach-avoid verification of stochastic discrete-time systems.
\newblock {\em Information and Computation\/} (2025), 105368.

\bibitem{yaghoubi2020risk}
{\sc Yaghoubi, S., Majd, K., Fainekos, G., Yamaguchi, T., Prokhorov, D., and Hoxha, B.}
\newblock Risk-bounded control using stochastic barrier functions.
\newblock {\em IEEE Control Systems Letters 5}, 5 (2020), 1831--1836.

\bibitem{vzikelic2023learning}
{\sc {\v{Z}}ikeli{\'c}, {\DJ}., Lechner, M., Henzinger, T.~A., and Chatterjee, K.}
\newblock Learning control policies for stochastic systems with reach-avoid guarantees.
\newblock In {\em AAAI Conf. on AI\/} (2023), vol.~37, pp.~11926--11935.

\end{thebibliography}


\appendix

\section*{Appendix}

\section{Benchmark Systems}
\label{appendixA}

\begin{table*}[t!]
\caption{\emph{Full} reach-avoid verification benchmark results for time-invariant certificates~\cite{mazouz2022safety}, one-shot time-varying certificates~\cite{mazouz2026stochastic}, and the proposed sequential time-varying certificates with uniform subspace regularization and initial-set tracking. Here, $\underline{P}_\reach$ is the certified lower bound on the reach-avoid probability,
$\hor$ the horizon, and \texttt{deg} the certificate degree. Further, $\tau$ is computation time in seconds. \texttt{OOM} denotes an out-of-memory failure and \texttt{TO} denotes a 1000s time-out. }
\label{tab:reach-avoid-verification-full}
\centering
\resizebox{0.80\textwidth}{!}{
\begin{tabular}{l|c|ccc|ccc|ccc|ccc}
\toprule
\multirow{2}{*}{Model}
& \multirow{2}{*}{$\hor$}
& \multicolumn{3}{c|}{\texttt{Time-Invariant}}
& \multicolumn{3}{c|}{\texttt{One-Shot TV}~\cite{mazouz2026stochastic}}
& \multicolumn{3}{c|}{\texttt{Sequential TV: Uniform}}
& \multicolumn{3}{c}{\texttt{Sequential TV: Tracking}}
\\
&
& \texttt{deg} & $\underline{P}_\reach$ & $\tau(s)$
& \texttt{deg} & $\underline{P}_\reach$ & $\tau(s)$
& \texttt{deg} & $\underline{P}_\reach$ & $\tau(s)$
& \texttt{deg} & $\underline{P}_\reach$ & $\tau(s)$
\\
\midrule

Linear 1D
& 10
& 4 & 0.581 & 0.04
& 2 & 0.837 & 0.32
& 4 & 0.805 & 0.11
& 4 & 0.712 & 0.08 \\
& 10
& 6 & 0.667 & 0.06
& 4 & 0.931 & 0.64
& 8 & 0.956 & 0.29
& 8 & 0.895 & 0.22 \\
& 10
& 8 & 0.714 & 0.08
& 6 & 0.981 & 1.23
& 12 & 0.985 & 0.87
& 12 & 0.972 & 0.65 \\
& 10
& 10 & 0.762 & 0.12
& 8 & 0.992 & 3.51
& 16 & 0.994 & 1.76
& 16 & 0.986 & 1.31 \\
& 10
& 12 & 0.795 & 0.17
& 10 & 0.996 & 9.58
& 20 & 0.997 & 3.89
& 20 & 0.995 & 2.94 \\
\cline{2-14}
& 20
& 6 & 0.686 & 0.07
& 4 & 0.952 & 1.49
& 8 & 0.965 & 0.59
& 8 & 0.906 & 0.44 \\
& 20
& 8 & 0.743 & 0.09
& 6 & 0.984 & 3.09
& 12 & 0.987 & 1.27
& 12 & 0.976 & 0.95 \\
& 20
& 10 & 0.776 & 0.13
& 8 & 0.994 & 6.91
& 16 & 0.996 & 2.51
& 16 & 0.988 & 1.88 \\
& 20
& 14 & 0.797 & 0.21
& 12 & 0.997 & 34.57
& 20 & 0.998 & 5.79
& 20 & 0.996 & 4.33 \\
\cline{2-14}
& 100
& 8 & 0.756 & 0.12
& 6 & 0.986 & 18.69
& 12 & 0.988 & 5.91
& 12 & 0.978 & 4.41 \\
& 100
& 10 & 0.791 & 0.18
& 8 & 0.995 & 54.76
& 16 & 0.997 & 11.48
& 16 & 0.989 & 8.61 \\
& 100
& 12 & 0.816 & 0.24
& 10 & 0.998 & 186.35
& 20 & 0.999 & 24.75
& 20 & 0.998 & 18.57 \\
& 100
& 14 & 0.823 & 0.32
& 12 & \textemdash & \texttt{OOM}
& 24 & 0.999 & 51.81
& 24 & 0.999 & 38.86 \\

\midrule

Contraction 2D
& 10
& 8 & 0.435 & 0.95
& 6 & 0.736 & 5.87
& 8 & 0.675 & 1.79
& 8 & 0.546 & 1.36 \\
& 10
& 10 & 0.491 & 1.51
& 8 & 0.816 & 10.31
& 12 & 0.827 & 4.22
& 12 & 0.752 & 3.15 \\
& 10
& 12 & 0.547 & 2.19
& 10 & 0.873 & 18.21
& 16 & 0.896 & 8.79
& 16 & 0.835 & 6.59 \\
& 10
& 16 & 0.662 & 3.87
& 12 & 0.916 & 36.55
& 20 & 0.936 & 17.71
& 20 & 0.892 & 13.27 \\
& 10
& 20 & 0.746 & 6.89
& 16 & 0.943 & 72.24
& 24 & 0.966 & 34.95
& 24 & 0.935 & 26.21 \\
& 10
& 24 & 0.753 & 12.53
& 20 & 0.965 & 318.68
& 28 & 0.985 & 72.51
& 28 & 0.967 & 54.39 \\
& 10
& 28 & 0.786 & 23.81
& 24 & \textemdash & \texttt{TO}
& 32 & 0.995 & 148.83
& 32 & 0.986 & 111.63 \\
\cline{2-14}
& 20
& 12 & 0.567 & 2.89
& 10 & 0.826 & 41.34
& 12 & 0.846 & 6.87
& 12 & 0.776 & 5.22 \\
& 20
& 16 & 0.652 & 5.19
& 12 & 0.876 & 96.62
& 16 & 0.916 & 13.51
& 16 & 0.856 & 10.18 \\
& 20
& 20 & 0.726 & 9.53
& 16 & 0.936 & 386.81
& 20 & 0.956 & 26.95
& 20 & 0.916 & 20.24 \\
& 20
& 24 & 0.766 & 17.91
& 20 & \textemdash & \texttt{TO}
& 24 & 0.976 & 45.82
& 24 & 0.956 & 34.23 \\
& 20
& 28 & 0.796 & 32.71
& 24 & \textemdash & \texttt{TO}
& 28 & 0.995 & 96.53
& 28 & 0.976 & 72.39 \\
\cline{2-14}
& 50
& 12 & 0.626 & 4.23
& 10 & 0.886 & 112.81
& 12 & 0.887 & 17.41
& 12 & 0.806 & 13.16 \\
& 50
& 16 & 0.706 & 7.89
& 12 & 0.936 & 284.71
& 16 & 0.937 & 35.52
& 16 & 0.886 & 26.67 \\
& 50
& 20 & 0.756 & 14.71
& 16 & 0.956 & 742.29
& 20 & 0.976 & 68.84
& 20 & 0.946 & 51.56 \\
& 50
& 24 & 0.776 & 26.93
& 20 & \textemdash & \texttt{TO}
& 24 & 0.986 & 114.73
& 24 & 0.966 & 85.52 \\
& 50
& 28 & 0.816 & 48.71
& 24 & \textemdash & \texttt{TO}
& 28 & 0.995 & 243.82
& 28 & 0.986 & 182.54 \\

\midrule

Aircraft 3D
& 10
& 4 & 0.000 & 0.84
& 2 & 0.316 & 8.79
& 4 & 0.256 & 3.01
& 4 & 0.136 & 2.25 \\
& 10
& 6 & 0.000 & 1.27
& 4 & 0.446 & 23.74
& 8 & 0.486 & 11.33
& 8 & 0.366 & 8.47 \\
& 10
& 8 & 0.000 & 2.18
& 6 & 0.556 & 59.11
& 12 & 0.616 & 38.81
& 12 & 0.516 & 29.12 \\
& 10
& 10 & 0.025 & 6.91
& 8 & 0.686 & 184.71
& 16 & 0.736 & 112.93
& 16 & 0.646 & 84.71 \\
& 10
& 12 & 0.066 & 18.51
& 10 & \textemdash & \texttt{OOM}
& 20 & 0.826 & 287.74
& 20 & 0.746 & 215.83 \\
\cline{2-14}
& 20
& 4 & 0.000 & 1.19
& 2 & 0.486 & 27.61
& 4 & 0.406 & 5.91
& 4 & 0.244 & 4.45 \\
& 20
& 6 & 0.000 & 1.86
& 4 & 0.676 & 89.55
& 8 & 0.746 & 22.63
& 8 & 0.596 & 16.91 \\
& 20
& 8 & 0.000 & 3.21
& 6 & 0.826 & 263.48
& 12 & 0.856 & 146.51
& 12 & 0.766 & 108.73 \\
& 20
& 10 & 0.036 & 9.83
& 8 & 0.886 & 716.65
& 16 & 0.916 & 326.84
& 16 & 0.846 & 244.94 \\
& 20
& 12 & 0.086 & 26.93
& 10 & \textemdash & \texttt{OOM}
& 20 & 0.956 & 684.65
& 20 & 0.906 & 513.53 \\
\cline{2-14}
& 30
& 4 & 0.000 & 1.58
& 2 & 0.536 & 43.69
& 4 & 0.456 & 8.87
& 4 & 0.274 & 6.68 \\
& 30
& 6 & 0.000 & 2.69
& 4 & 0.716 & 141.47
& 8 & 0.776 & 34.15
& 8 & 0.636 & 25.55 \\
& 30
& 8 & 0.016 & 5.18
& 6 & 0.846 & 394.82
& 12 & 0.886 & 219.85
& 12 & 0.806 & 163.32 \\
& 30
& 10 & 0.056 & 14.91
& 8 & 0.906 & 864.38
& 16 & 0.946 & 492.74
& 16 & 0.886 & 369.58 \\

\midrule

Room Temp.\ 2D
& 10
& 4 & 0.126 & 4.89
& 2 & 0.516 & 51.94
& 4 & 0.467 & 11.49
& 4 & 0.367 & 8.77 \\
& 10
& 6 & 0.226 & 9.53
& 4 & 0.636 & 96.51
& 8 & 0.686 & 35.51
& 8 & 0.576 & 26.91 \\
& 10
& 8 & 0.316 & 18.81
& 6 & 0.726 & 194.68
& 12 & 0.796 & 91.83
& 12 & 0.706 & 69.27 \\
& 10
& 10 & 0.444 & 36.93
& 8 & 0.816 & 453.23
& 16 & 0.906 & 208.84
& 16 & 0.836 & 157.53 \\
& 10
& 12 & 0.516 & 72.57
& 10 & 0.856 & 824.47
& 20 & 0.946 & 426.94
& 20 & 0.896 & 320.53 \\
& 10
& 14 & 0.576 & 138.73
& 12 & \textemdash & \texttt{TO}
& 24 & 0.976 & 786.53
& 24 & 0.946 & 589.92 \\
\cline{2-14}
& 50
& 4 & 0.386 & 6.91
& 2 & 0.656 & 168.45
& 4 & 0.597 & 38.73
& 4 & 0.477 & 29.36 \\
& 50
& 6 & 0.536 & 13.91
& 4 & 0.766 & 326.95
& 8 & 0.823 & 146.84
& 8 & 0.716 & 111.03 \\
& 50
& 8 & 0.646 & 28.57
& 6 & 0.866 & 714.37
& 12 & 0.906 & 367.95
& 12 & 0.836 & 276.54 \\
& 50
& 10 & 0.724 & 56.83
& 8 & \textemdash & \texttt{OOM}
& 16 & 0.946 & 692.97
& 16 & 0.896 & 519.76 \\

\midrule

Room Temp.\ 3D
& 10
& 2 & 0.000 & 42.91
& 2 & 0.246 & 428.81
& 2 & 0.204 & 162.93
& 2 & 0.156 & 128.55 \\
& 10
& 4 & 0.000 & 118.73
& 4 & \textemdash & \texttt{TO}
& 4 & 0.467 & 684.84
& 4 & 0.396 & 517.53 \\
& 10
& 6 & 0.000 & 326.95
& 6 & \textemdash & \texttt{TO}
& 6 & 0.716 & 926.94
& 6 & 0.636 & 914.73 \\

\bottomrule
\end{tabular}
}
\end{table*}

\subsection{Safety Systems}

\paragraph{Linear Systems}
To showcase the intuition of each framework, we consider two linear stochastic systems with stable ($\px_{k+1} = 0.95 \px_k + \pw_k)$  and unstable ($\px_{k+1} = 1.05 \px_k + \pw_k)$ dynamics and additive noise \( \pw_k \sim \N(0, 0.01) \).
The systems operate in 
\( X_{\safe} = [-1.0, 0.5]^2 \) and \( X_{\initial} = [\pm 0.05]^2 \). 

\paragraph{Oscillator}
The Van der Pol oscillator from~\cite{abate2020arch} is modeled as a discrete-time stochastic system with dynamics 
%
\begin{align*}
    \px^{(1)}_{k+1} &= \px^{(1)}_k + \lambda \px^{(2)}_k + \pw^{(1)}, \\
    \px^{(2)}_{k+1} &= \px^{(2)}_k + \lambda\big(-\px^{(1)}_k + (1 - (\px^{(1)}_k)^2) \, \px^{(2)}_k\big) + \pw^{(2)},
\end{align*}
where \( \lambda= 0.1 \) is the sampling time and \( \pw_k \sim \N(0, 4 \cdot 10^{-4} I) \).
%
Further, \( X = [-7, 7]^2 \), and the initial set is \( X_{\initial} = [-5, 5]^2 \).

\paragraph{Lokta-Volterra}

For the third benchmark, we consider a discrete-time Lotka-Volterra type stochastic system with state \( \px_k = (v_k, p_k) \in \mathbb{R}^2 \), where \( v \) and \( p \) denote the prey and predator populations, respectively. The dynamics are 
\begin{align*}
v_{k+1} &= v_k + T \left(\theta v_k (1 - v_k) - \phi v_k p_k \right) + \pw_1, \\
p_{k+1} &= p_k - T \left(\psi p_k - \delta v_k p_k \right) + \pw_2,
\end{align*}
where \( T = 0.1 \), \( \theta = 1.1 \), \( \phi = 0.4 \), \( \psi = 0.4 \), and \( \delta = 0.1 \). The noise satisfies \( \pw_k \sim \N(0, \mathrm{diag}(0.01,\,0.005)) \).

The state space is \( X = [0, 10] \times [0, 5] \), and the initial set is \( X_{\initial} = [6, 7] \times [2, 3] \).

\paragraph{Dubin's Car}
For the fourth benchmark, we consider a Dubin's car model with a polynomial representation of the trigonometric dynamics. The state is given by \( \px_k = (p_x(k), p_y(k), \theta_k, v_k) \in \mathbb{R}^4 \), where \( (p_x, p_y) \) denote position, \( \theta \) is the heading angle, and \( v \) is the speed. The dynamics are 
\begin{align*}
p_x(k+1) &= p_x(k) + T \, v_k \left(1 - \tfrac{1}{2}\theta_k^2 \right) + \pw_1, \\
p_y(k+1) &= p_y(k) + T \, v_k \left(\theta_k - \tfrac{1}{6}\theta_k^3 \right) + \pw_2, \\
\theta_{k+1} &= \theta_k + T \, \omega + \pw_3, \\
v_{k+1} &= v_k + \pw_4,
\end{align*}
where \( T = 0.05 \) is the sampling time, \( \omega = 0.8 \) is a constant turn rate, and the noise satisfies \( \pw_k \sim \N(0, \mathrm{diag}(0.05,\,0.05,\,0.02,\,0.02)) \).

The state space is \( X = [-5,5] \times [-5,5] \times [-1.5,1.5] \times [0,3] \), and the initial set is \( X_{\initial} = [-0.5,0.5] \times [-0.5,0.5] \times [-0.2,0.2] \times [0.5,1.0] \). 

\paragraph{Quadrotor 6D}

We finally consider a quadrotor guidance model consisting of lateral and vertical dynamics, where the longitudinal and spin variables are constrained. By fixing the vertical and spin variables, we obtain the lateral and longitudinal motion model. The resulting closed-loop state space model is given by 
\begin{align*}
    &\Delta \dot{y}_{E} = \Delta v, &&
    \Delta \dot{v} = g \Delta \phi, \\
    & \Delta \dot{\phi} = \Delta p, &&
    \Delta \dot{p} = \frac{1}{I_x} \Delta L_c,
\end{align*}
where $g$ denotes the gravitational acceleration and $I_x$ is the moment of inertia about the $x$-axis. The state $y_{E} \in [-0.5, 2.0]$ represents the lateral position, while $v \in [-1.0, 1.0]$ is the corresponding velocity. The roll angle satisfies $\phi \in [-0.1, 0.1]$, the roll rate satisfies $p \in [-0.1, 0.1]$, and $L_c$ denotes the roll control moment. The complete dynamics constitute a six-dimensional linear model. Using a sampling time of $\Delta t = 0.01$, we discretized the system using the forward Euler method and added process noise $\mathbf{v} \sim \mathcal{N}(0, 10^{-4}I)$.

\subsection{Reach-Avoid Systems}

\paragraph{1D Linear Systems}
To showcase the intuition of each approach, we consider two one-dimensional systems with stable and unstable dynamics, both with $\pw_k \sim \N(0, 0.1)$. The dynamics of the first system are
\begin{align}
    \label{eq:1D_linear_stable}
    \px_{k+1} = 0.75 \px_k + \pw_k,
\end{align}
with \( X_{\safe} = [1, 3] \), \( X_{\initial} = [1.8, 1.9] \), \( X_{\reach} = [0.7, 1.0] \), and \( X_{\unsafe} = [3.0, 3.3] \).
The second system is
\begin{align}
    \label{eq:1D_linear_unstable}
    \px_{k+1} = 1.15 \px_k + \pw_k,
\end{align}
with \( X_{\safe}  = [1, 3] \), \( X_{\initial} = [2.2, 2.3] \),
\( X_{\reach} = [3.0, 3.5] \), and \(  X_{\unsafe}= [0.7, 1.0] \). 
For both the above systems, the control $U \in [-0.25, 0.25]$ for linear feedback control synthesis.

\paragraph{2D Contraction Maps}

Next, we consider two \emph{contraction} systems in $\reals^{2}$ with a rotation. For the first contraction map, the dynamics are given as
\begin{align}
    \label{eq:contraction_map_1}
    \px_{k+1} =
    \begin{bmatrix}
        0.75 & 0.15\\
        -0.15 & 0.50
    \end{bmatrix}\px_k + \pw_k,
\end{align}
with $X_{\safe} = [-0.2, 2]\times[-0.5, 0.5],~X_{\initial} = [1.6, 1.7]\times[-0.1, 0.1],$, $X_{\reach} = [-0.2, 0.2]\times[-0.5, 0.5],$ and the unsafe region is around the safe set.  The noise is $\pw_k \sim \N(0, 10^{-2} I)$. 

The second case study has dynamics
\begin{align}
    \label{eq:contraction_map_2}
    \px_{k+1} =
    \begin{bmatrix}
        0.50 & 0.55\\
        -0.35 & 0.50
    \end{bmatrix}\px_k + \pw_k,
\end{align}
with the setup being the same as the previous case.

\paragraph{3D Aircraft Model}
\label{sec:aicraft}
We also consider a linearized aircraft model in $\reals^{3}$, with continuous-time longitudinal dynamics
{\footnotesize
\begin{align*}
    \dot{\px} =
    \begin{bmatrix}
        X_u/m & X_w/m & -g\cos\theta_0/m\\
        Z_u/(m-Z_{\dot w}) & Z_w/(m-Z_{\dot w}) & (Z_q+mU_0)/(m-Z_{\dot w})\\
        M_u/I_{yy} & M_w/I_{yy} & M_q/I_{yy}
    \end{bmatrix} \px,
\end{align*}
}
where $\mathbf{x} = [u, w, q]^\top$ denotes the forward velocity, vertical velocity, and pitch rate perturbations. Further, $m$ denotes the aircraft mass, and $I_{yy}$ the pitch-axis moment of inertia. We used a forward Euler scheme with $\Delta t = 0.1$ to discretize time, and use a LQR controller to stabilize the system near the reach set. Then noise $\pw_k \sim \N(0, 10^{-2}I)$ was added to the system. 
All parameter values of this model are provided in~\cite{mazouz2026timevarying}.

\paragraph{2D \& 3D Room Temperature}
We finally consider a 2D and a 3D room temperature system with polynomial dynamics.  The 2D model is
{\small
\begin{align*}
    \textbf{x}_1(k+1) & = (1 - \tau(\alpha + \alpha_e)) \px_1(k) + \tau \alpha \px_2(k) + \tau \alpha_e T_e + \pw_1(k) \\
    \textbf{x}_2(k+1) & = (1 - \tau(2\alpha + \alpha_e)) \px_2(k) + \tau \alpha \px_1(k) + \tau \alpha_e T_e + \pw_2(k)
\end{align*}
}
where $\pw \sim \N(0, 10^{-2}I)$. The states $\px_1$ and $\px_2$ define the temperature of each room. The ambient temperature $T_e = 10^\circ$C, while the heat coefficients are $\alpha_e = 8 \times 10^{-3}$ and $\alpha = 6.2 \times 10^{-3}$. The sampling time $\tau$ = 5 minutes. 

\paragraph{Van der Pol 6D}

The six-dimensional polynomial Van der Pol system is constructed from three independent Van der Pol oscillators. The dynamics are given by
\begin{align*}
    \dot{x}_1 &= x_2, \\
    \dot{x}_2 &= \mu x_2-\mu x_1^2x_2-x_1, \\
    \dot{x}_3 &= x_4, \\
    \dot{x}_4 &= \mu x_4-\mu x_3^2x_4-x_3, \\
    \dot{x}_5 &= x_6, \\
    \dot{x}_6 &= \mu x_6-\mu x_5^2x_6-x_5,
\end{align*}
where $\mu>0$ determines the strength of the nonlinear damping. The resulting vector field is polynomial, with nonlinear terms of degree three. The state is $x=(x_1,\ldots,x_6)\in\mathbb{R}^6$, where each pair of states describes the position and velocity of an oscillator. Each oscillator exhibits a stable limit cycle, making the system a useful benchmark for evaluating nonlinear analysis and reachability methods.

\paragraph{Quadrotor 8D}

Following the same approach as for the lateral guidance, the longitudinal guidance is designed based on the dynamics
\begin{align*}
& \Delta \dot{x}_{E} = \Delta u, &&
\Delta \dot{u} = -g \Delta \delta, \
& \Delta \dot{\theta} = \Delta q, &&
\Delta \dot{q}= \frac{1}{I_y} \Delta M_c,
\end{align*}
where $g$ denotes the gravitational acceleration and $I_y$ is the moment of inertia about the $y$-axis. The state variables are the $x$-position $x_E \in [-0.5, 4.0]$, the corresponding velocity $u \in [-0.5, 1.5]$, the pitch angle $\theta \in [-0.1, 0.1]$, and the pitch rate $q \in [-0.1, 0.1]$. Here, $M_c$ denotes the pitch control moment. The continuous-time dynamics are discretized using the forward Euler method with a sampling period of $\Delta t = 0.01$, with additive Gaussian process noise $\mathbf{v} \sim \mathcal{N}(0,10^{-4}I)$.

\section{Full Results}
\label{app:full}

The full reach-avoid verification results of Tab.~\ref{tab:reach-avoid-verification-highlight} are reported in Tab.~\ref{tab:reach-avoid-verification-full}. These results further illustrate the scalability advantages of sequential synthesis across polynomial degrees and horizons, enabling tighter certified probability bounds where the oneshot formulation encounters computational limitations.

\clearpage
\thispagestyle{empty}

\end{document}